\documentclass[aps,pra,twocolumn,showpacs,preprintnumbers,amsmath,amssymb,superscriptaddress,longbibliography]{revtex4-1}
\pdfoutput=1

\usepackage{mathrsfs}
\usepackage{amsfonts}
\usepackage{amsmath}
\usepackage{amssymb}
\usepackage{natbib}
\usepackage{bbm}
\usepackage{xcolor}

\usepackage{graphicx}% Include figure files

\begin{document}

%\preprint{APS/123-QED}

\title{Faraday imaging induced squeezing of a double-well Bose-Einstein condensate}% Force line breaks with \\
%\title{Optical generation of spin squeezed state in a double-well trap}
%\title{Optical generation of spin squeezed state using atom Bose-Einstein condensate in a double-well trap}

\author{Ebubechukwu O. Ilo-Okeke}
%\author{Author List}
\affiliation{New York University Shanghai, 1555 Century Ave, Pudong, Shanghai 200122, China}  
%\affiliation{Department of Physics, School of Physical Sciences, Federal University of Technology, P. M. B. 1526, Owerri 460001, Nigeria}

\author{Shinichi Sunami}
\affiliation{Clarendon Laboratory, University of Oxford, Oxford OX1 3PU, United Kingdom}

\author{Christopher J. Foot}
\affiliation{Clarendon Laboratory, University of Oxford, Oxford OX1 3PU, United Kingdom}

\author{Tim Byrnes}
\affiliation{State Key Laboratory of Precision Spectroscopy, School of Physical and Material Sciences,
East China Normal University, Shanghai 200062, China}
\affiliation{New York University Shanghai, 1555 Century Ave, Pudong, Shanghai 200122, China}
\affiliation{NYU-ECNU Institute of Physics at NYU Shanghai, 3663 Zhongshan Road North, Shanghai 200062, China}
\affiliation{National Institute of Informatics, 2-1-2 Hitotsubashi, Chiyoda-ku, Tokyo 101-8430, Japan}
\affiliation{Department of Physics, New York University, New York, NY 10003, USA}

\date{\today}% It is always \today, today,

%%%%%%%%%%%%%%%%%%%%%%%%%%%%%%%%%%%%%%%%%%%%%%%%%%%%%%%%%%%%%%%%%%
%%%%%%%%%%%%%%%%%%%%%%%%%%%%%%%%%%%%%%%%%%%%%%%%%%%%%%%%%%%%%%%%%%
\begin{abstract}
We examine how non-destructive measurements generate  spin squeezing in an atomic Bose-Einstein condensate confined in a double-well trap. The condensate in each well is monitored using coherent light beams in a Mach-Zehnder configuration that interacts with the atoms through a quantum nondemolition Hamiltonian. We solve the dynamics of the light-atom system using an exact wavefunction approach, in the presence of dephasing noise, which allows us to examine arbitrary interaction times and a general initial state. We find that monitoring the condensate at zero detection current and with identical coherent light beams minimizes the backaction of the measurement on the atoms. In the weak atom-light interaction regime, we find the mean spin direction is relatively unaffected, while the variance of the spins is squeezed along the axis coupled to the light. Additionally, squeezing  persists in the presence of tunneling and dephasing noise.
\end{abstract}

%\pacs{}% PACS, the Physics and Astronomy
\maketitle

%%%==========================================================
%%% Introduction
%%%==========================================================
\section{Introduction \label{sec:intro}}
%%%==========================================================
Squeezed states of quantum systems are a resource that have numerous applications. They possess entanglement that can be used advantageously in various  quantum information applications~\cite{nielsen2000}, quantum metrology~\cite{giovannetti2011,degen2017}, precision measurements~\cite{appel2009,dariano2001}, time keeping~\cite{jozsa2000,louchet-chauvet2010,komar2014,ilo-okeke2018}, and quantum networks~\cite{elliot2002,hahn2019,ilo-okeke2020,nagele2020}. Squeezed states~\cite{giovannetti2011,kitagawa1993,ma2011} give phase sensitivities that beat the standard quantum limit~\cite{wineland1992,kitagawa1993} set by the quantum noise of individual uncorrelated quantum particles~\cite{wineland1994}. Consequently, the use of entanglement and squeezing for enhanced measurement and detection are becoming increasingly common in many areas of physics such as image reconstruction~\cite{brida2010}, magnetometry and electric field sensing~\cite{brask2015,fan2015,degen2017}, optical interferometry~\cite{dowling2015,schnabel2017}, and gravitational wave detection~\cite{eberle2010,pitkin2011}. 

Realizing squeezed states in different systems is typically dependent on generating quantum correlations or entanglement created by the nonlinear interactions between the quantum particles. For instance, the Kerr effect~\cite{boyd2008,new2011} is largely exploited in creating optical squeezed states~\cite{slusher1985,wu1986}, and  has been widely studied~\cite{scully1997,loudon2000,gerry2005} and applied in various technologies~\cite{dowling2015,schnabel2017,dowling2008,joana2016steady,ishida2013photoluminescence}. In atomic systems such as Bose-Einstein condensates (BECs)~\cite{kitagawa1993,choi2005,esteve2008,bohi2009,jing2019} and trapped ions~\cite{leibfried2004,ge2019}, two-body spin interactions~\cite{kitagawa1993} have been widely exploited for generation of entanglement and squeezing.

Techniques such as quantum nondemolition (QND) measurement~\cite{takahashi1999,higbie2005,kuzmich2004,meppelink2010,ilo-okeke2014,ilo-okeke2016} have also been used to realize squeezing~\cite{appel2009,schleier-smith2010,sewell2012,cox2016,hosten2016} in atom systems. In this approach, the phase shift acquired by the light pulse after passing through the atomic samples is measured. The light probe has a large frequency detuning from the atomic resonance transition to give very low photon scattering rates making the measurement minimally-destructive. Another advantage of the measurement technique is that the information about the atomic population is contained in the phase of light making it available for collection and readout. These features are utilized in minimally destructive measurement of atom systems to produce entanglement between hyperfine levels of atoms~\cite{behbood2014,kong2020}, and observe spin rotations of atoms placed in an rf field below the standard quantum limit~\cite{kuzmich2000}. 

Recently, Vasilakis, M{\o}ller, Polzik, and co-workers used a stroboscopic back-action-evading measurement~\cite{vasilakis2015,moller2017} to generate a squeezed state of an oscillator. The oscillator in this case consisted of a polarized atomic ensemble. On applying a magnetic field parallel to the polarization axis, the spin undergoes precession about the applied field. A  probe applied to the plane perpendicular to the magnetic field axis couples to one of the spin components on that plane. By manipulating the strength of atom-probe interactions, one can increase the correlations between the atom and light. It is found that the times where the oscillator strength is maximum, the noise of the measurement device is not coupled to the oscillator~\cite{thorne1978,Braginsky1980}. Hence for measurements performed at these times, the normalized variance  of the atom operator appearing in the atom-light interaction Hamiltonian is squeezed.

\begin{figure}[t]
	\includegraphics[width=\columnwidth]{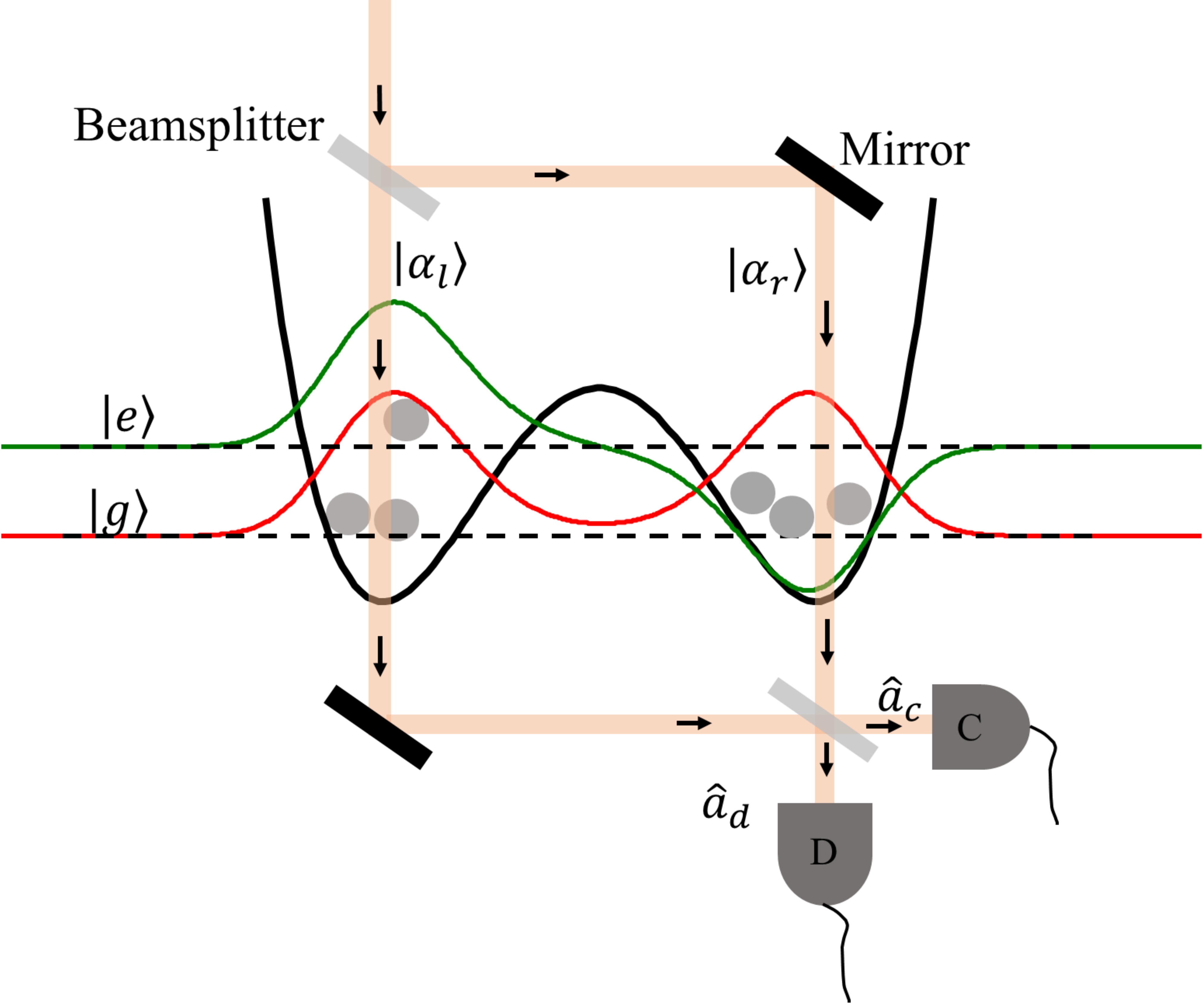}
	\caption{A double-well trap containing atoms that are allowed to tunnel between the wells. The atoms in each well are imaged using identical laser light in the Mach-Zehnder interferometer configuration. The solid black line is the double-well trap. The symmetric wavefunction is the ground state $\lvert g\rangle$ wavefunction and the antisymmetric line is first excited state $\lvert e\rangle$ belonging to the double-well trap, respectively.}
	\label{fig1}
\end{figure}

Here we propose using the balanced detection scheme in a Mach-Zehnder interferometer to generate a squeezed state of an oscillator as shown in Fig~\ref{fig1}. This technique does not rely on performing measurements at the times where the oscillator strength is maximum in order to generate squeezing. Rather, it requires the optical coherent state of light in the two arms of the interferometer to have the same amplitude and phase $\lvert \alpha_l\rangle = \lvert \alpha_r\rangle$. For the case of zero detection current ($n_d = n_c$), the measurement back-action is minimized, and atom-light interactions modulate the BEC state. For sufficiently strong  atom-light coupling strength, squeezing appears. 

To describe how the squeezing comes about, we consider an atomic BEC that is prepared in the ground state of a double-well potential~\cite{milburn1997}, as shown in Fig~\ref{fig1}. As such, atoms are completely delocalized between the two wells.  Working in a two-level approximation of the double-well trap, where the eigenstates are symmetric and antisymmetric wavefunctions, we may form a pseudospin. To produce squeezing, laser light is used to perform an interferometric  measurement on the atoms, as shown in Fig.~\ref{fig1}, under the conditions specified above. The measurement of the photons induces a modification on the quantum state of the spins, while leaving its mean spin direction unchanged. This is because the photon detection causes a random evolution of atoms such that on the average the net motion of the mean spin projection on the plane perpendicular to the mean spin direction is zero, thereby leaving the mean spin direction unchanged. Additionally, the photon detection constricts the random motion of the mean spin direction along the spin axis that is coupled to light. The net effect is that the random motion of the atoms is only able to shear the initial distribution in phase space, with the resulting distribution having a reduced variance. The reduced variance of the distribution caused by the detection of photons is the squeezing, and  manifests as a decrease in variance below $\sqrt{N}$ for the spin component that is coupled to light.

We investigate and characterize the squeezed spin state generated in the double-well trap using a QND measurement as described above. In the zero tunneling limit, we find a simple analytic expression for the probability density that confirms the spin state after photon detection is indeed squeezed. We develop a more realistic model that includes both tunneling and dephasing noise from the measurement and show that squeezing persists in the presence of dephasing noise. One of the main features of our work is that we use an exact wavefunction method to solve the dynamics of the quantum nondemolition Hamiltonian.  In many works relating to light generated squeezing, the Holstein-Primakoff approximation is used for the atomic spins~\cite{hald1999,julsgaard2001,vasilakis2015,moller2017}.  This is valid for highly spin polarized initial spins, and for short evolution times, but at longer evolution times it loses validity~\cite{tim2015}.  Our methods do not have this restriction and can be applied in a more general setting. Additionally, experiments with stroboscopic measurement~\cite{vasilakis2015,sewell2012,moller2017} require exquisite engineering, precision and control in order to reduce the noise of the pulsed laser light to below the required level, whereas  the continuous measurement as analysed here has less experimental parameters that could introduce imperfection and noise.

The remainder of the paper is organized as follows. We start by developing the theoretical framework for describing the tunneling of atoms in a double-well potential and subsequent squeezing using the QND measurement, as well as describing the framework for dephasing of the BEC in Sec.~\ref{sec:model}. We develop an exact wavefunction approach to explaining the origin of squeezing due to QND measurement in Sec.~\ref{sec:simplemodel}, and study the effect of tunneling on the squeezed state in Sec.~\ref{sec:blochspheretunneling}. Next in Sec.~\ref{sec:blochspheredephasing}, we make the model more realistic by including the effects of dephasing on the squeezed state of atoms in the double-well trap.  We discuss how this scheme can be realized with existing experimental technology, in Sec.~\ref{sec:experiments}. Finally, the summary and conclusions are presented in Sec.~\ref{sec:discussion}.

%%===========================================
%%Figure 1
%%===========================================
%%===========================================

%%%==========================================================
\section{Double-well trap model and measurement of atoms}\label{sec:model}
%%%==========================================================
\subsection{Tunneling in a double-well trap}
%%%==========================================================
Our model system is a many-body atomic Bose-Einstein condensate confined in a double-well potential that can be made from magnetic~\cite{schumm2005,harte2018} or optical~\cite{shin2004}  fields. This is illustrated in Fig.~\ref{fig1}. The many-body Hamiltonian governing the dynamics of the atoms in double-well potential $V(r)$ is  
\begin{align}	
\label{eq:strn01}
\hat{H}& =\int d\mathbf{r} \hat{\Psi}^\dagger(\mathbf{r},t)\left[\frac{P^2}{2M} + V(\mathbf{r})\right]\hat{\Psi}(\mathbf{r},t)  \nonumber\\ 	 
&+ \frac{U_0}{2}\int d\mathbf{r}\hat{\Psi}^\dagger(\mathbf{r},t) \left(\hat{\Psi}(\mathbf{r},t) \hat{\Psi}^\dagger(\mathbf{r},t) - 1\right)\hat{\Psi}(\mathbf{r},t),
\end{align}
where $U_0  = 4\pi\hbar^2a_\mathrm{s}/M$ is the two-body contact interaction energy, $a_s$ is the s-wave scattering length, $M$ is the mass of an atom, and $P$ is the momentum of an atom. 

Assuming that the fluctuation about the mean atom number density $\langle n(\mathbf{r}) \rangle$ is small,  we can perform a mean-field approximation and neglect the quadratic noise terms. We approximate the contact interaction Hamiltonian as 
\begin{align}
\label{eq:papc02}
\hat{H}_\mathrm{int} & = \frac{U_0}{2} \int d\mathbf{r} \bigg(- \langle n(\mathbf{r})\rangle^2 + \hat{\Psi}^\dagger(\mathbf{r},t) \Big[2\langle n(\mathbf{r}) \rangle \nonumber\\
& -1)\Big] \hat{\Psi}(\mathbf{r},t)\bigg),
\end{align}
where the first term merely adds a constant energy term, which can be ignored  so that the Hamiltonian of the system (\ref{eq:strn01}) becomes
\begin{align}
\label{eq:papc03}
\hat{H} & =\int d\mathbf{r} \hat{\Psi}^\dagger(\mathbf{r},t)\left[\frac{P^2}{2M} + V(\mathbf{r}) + U_0 \langle n(\mathbf{r}) \rangle \right]\hat{\Psi}(\mathbf{r},t).
\end{align}
This single-particle Hamiltonian is of  Gross-Pitaevskii form~\cite{dalfovo1999,leggett2001}. 

We assume that only two states are allowed in the trap. Solving the Gross-Pitaevskii equation for the double-well potential, we find the ground state $\psi_{g}(\mathbf{r},t)$ is a symmetric state  with energy $\hbar\omega_{g}$ and the first excited state $\psi_{e}(\mathbf{r},t)$ is the antisymmetric state with energy $\hbar\omega_{e}$. Let the  field operator $\hat{\Psi}(\mathbf{r},t)$ be expanded  in terms of the ground and excited solutions, $\hat{\Psi}(\mathbf{r},t) = \hat{b}_g\psi_\mathrm{g}(\mathbf{r},t) + \hat{b}_e\psi_\mathrm{e}(\mathbf{r},t)$, where $\hat{b}_{g,e}$ is an annihilation operator. This acting on the vacuum destroys it, $\hat{b}_{g,e} \lvert0\rangle= 0$. Due to orthogonality conditions, the overlap vanishes, $\int d\mathbf{r} \psi^*_e(\mathbf{r},t)\psi_g(\mathbf{r},t) = 0$, and the Hamiltonian density does not couple the ground and the excited states of the macroscopic oscillator. Substituting for the field operator in~(\ref{eq:papc03}) gives 
\begin{align}
\label{eq:strn02}
\hat{H} &=  \frac{\hbar}{2}\left(\omega_e + \omega_g \right)\hat{N} +\hbar\Omega J_z,
\end{align}
where $\hat{N} = \hat{b}^\dagger_g \hat{b}_g + \hat{b}^\dagger_e \hat{b}_e$ is the total particle number operator, $J_z = (\hat{b}^\dagger_e \hat{b}_e - \hat{b}^\dagger_g \hat{b}_g)/2$ measures the relative atom number in both the ground and excited states, and $\Omega = \omega_{e} - \omega_{g}$ is the particle tunneling frequency. The other spin operators are $J_x = (\hat{b}^\dagger_e \hat{b}_g + \hat{b}^\dagger_g \hat{b}_e)/2$, $J_y = -i( \hat{b}^\dagger_e \hat{b}_g - \hat{b}^\dagger_g \hat{b}_e)/2$. In the following, we will neglect the constant energy term in (\ref{eq:strn02}) as it contributes only a shift in the energy level.

The initial state of atoms is taken to be in a linear superposition of symmetric and antisymmetric states as 
\begin{equation}
\label{eq:strn07}
\lvert \alpha,\beta\rangle =  \frac{1}{\sqrt{N!}}\left( \alpha \hat{b}_e^\dagger + \beta \hat{b}_g^\dagger  \right)^N\lvert 0 \rangle,
\end{equation}
where $\lvert 0\rangle$ is the vacuum state for the atom. The probability $|\alpha|^2$ gives the fractional population of atoms found in the excited state of the double-well potential, while  $|\beta|^2$ gives the fractional number of atoms found in the ground state of the double-well potential. These probabilities satisfy $|\alpha|^2 + |\beta|^2 = 1$. We may also write (\ref{eq:strn07}) in terms of Bloch sphere angles~\cite{arecchi1972,gross2012},
\begin{equation}
	\label{eq:strn:me06}
	\lvert\theta,\phi\rangle =  \frac{1}{\sqrt{N!}}\left( \sin(\theta/2) e^{-i\phi/2} \hat{b}_e^\dagger + \cos(\theta/2) e^{i\phi/2} \hat{b}_g^\dagger  \right)^N\lvert 0 \rangle.
\end{equation}
%for $\alpha = \sin(\theta/2) e^{-i\phi/2}$ and $\beta = \cos(\theta/2) e^{i\phi/2}$. 

%==================================================
\subsection{Light-atom interaction}
%==================================================
In order to discuss the physics of the double-well it is convenient to consider atoms localized in the left and right well, using the  transformation $\hat{b}_g = (\hat{b}_l + \hat{b}_r)/\sqrt{2}$ and $\hat{b}_e = (\hat{b}_l - \hat{b}_r)/\sqrt{2}$. The atom operators $J_x$, $J_y$ and $J_z$ defined in terms of the operators $b_l$ and $b_r$ become
\begin{align}
\label{eq:pap01}
J_x & = \frac{\hat{b}_l^\dagger \hat{b}_l - \hat{b}^\dagger_r \hat{b}_r}{2}\nonumber,\\
J_y & = \frac{\hat{b}^\dagger_l \hat{b}_r - \hat{b}^\dagger_r \hat{b}_l}{2i},\\
J_z & = \frac{b^\dagger_l \hat{b}_r + \hat{b}^\dagger_r \hat{b}_l}{2}.\nonumber
\end{align} 
The total number operator $\hat{N} = \hat{b}^\dagger_l \hat{b}_l + \hat{b}^\dagger_r \hat{b}_r$, is a conserved quantity. In this basis, the operator $J_x$ gives the relative atom number in each well. The operator $J_z$ gives the relative atom number difference between the excited and  ground states of the double-well trap. For instance $\langle J_z\rangle$ being positive implies that there are more atoms in the ground state of the double-well trap than in the excited state and vice versa. Finally, $J_y$ gives the relative phase  between the condensates in the two wells.

We consider atoms in the double-well potential interacting with laser light that is detuned from the atomic resonance transition~\cite{higbie2005,meppelink2010}, as shown in Fig.~\ref{fig1}. This set up  can be described by the QND Hamiltonian~\cite{julsgaard2001,kuzmich2004,behbood2014,ilo-okeke2014}. We write the total Hamiltonian of the system including the light-matter interaction as 
\begin{equation}
\label{eq:strn11}
H_\mathrm{eff} = \hbar\Omega J_z +  2\hbar g J_x S_z,
\end{equation}
where  $g $ is the atom-light coupling strength~\citep{ilo-okeke2014}. Writing the sum of atom-light interactions in both wells in terms of the relative photon number $S_z =(\hat{a}^\dagger_l \hat{a}_l - \hat{a}^\dagger_r\hat{a}_r)/2$ that passes through the wells gives rise to the dependence of atom-light interactions on the spins. The total number operator that introduces an overall global phase has been ignored in writing $H_\mathrm{eff}$.

%%%==========================================================
\subsection{State evolution in the presence of ac Stark shift dephasing\label{sec:sec:dephasing}}
%%%==========================================================  
Quantum states of matter such as the spin squeezed state that we generate are usually fragile and are susceptible to decoherence~\cite{lone2015} and particle loss~\cite{li2008}  caused by  interactions with the  environment. For example, this may include imperfections in the experiment, interactions among the particles, and the interactions with the probe. The statistical fluctuations resulting from  these interactions introduces an inherent randomness in the quantum state of particles that interacts with the environment. The randomness introduced by the environment leads to  the randomization of the phase of the particles that  makes it impossible to recover the coherence properties of the quantum state, even with postselection. For example, photon scattering by atoms, which leads to a decay in atomic polarization~\cite{grimm2000}, plays a role in the light shift dephasing of BEC state. This is because the finite line width used in calculating the ac Stark shift or light shift is also relevant in determining the scattering rate. As such, the atomic operator $J_x$ coupling the atoms to light as in (\ref{eq:strn11}) creates an ac Stark dephasing channel for the atoms. This allows us to describe the effective dynamics of the coupled atom-light system by a Markovian master equation~\cite{byrnes2013,lone2015}  for the density operator $\rho$
\begin{align}
\label{eq:strn:me01}
\frac{d\rho}{dt} = -\frac{i}{\hbar}\left[H_\mathrm{eff},\rho\right]
+ \gamma\left( J_x\rho J_x - \frac{1}{2}\left\{ J_x^2, \rho\right\} \right),
\end{align}
where $\gamma$ is the dephasing rate, and $H_\mathrm{eff}$ is as given in (\ref{eq:strn11}).

At the end of the evolution of the density matrix, the light beams leaving the trap are recombined as shown in Fig.~\ref{fig1}. These photons are detected at the detector $c$ and $d$, respectively, thereby collapsing the coherent state of light. The full set of equations used to solve the master equation is given in Appendix \ref{sec:masterequation}. One of the techniques that we use is to expand the photons in the coherent state basis, this avoids an expansion in the Fock basis which is much more computationally demanding.  This is possible because the Hamiltonian (\ref{eq:strn11}) only couples to the light with photon number operators, which evolve the coherent states by applying a phase. The state of the atomic BEC given that $n_c$ and $n_d$ photons have been detected is $\rho_{n_c,n_d}$ (\ref{eq:strn:me10}). From here it is straightforward to calculate the averages of BEC operators as $\langle A \rangle  = \mathrm{Tr}\left[A\rho_{n_c,n_d}\right]$. For instance, the relative number of atoms in the wells given that $n_c$ and $n_d$ has been detected may be inferred by the expectation value of $J_x$ defined as $\langle J_x \rangle_{n_c, n_d} = \mathrm{Tr}\left[J_x\rho_{n_c,n_d}\right]$. Similarly, the variance is defined as 
\begin{equation}
	\label{eq:pap02}
	\left(\Delta J_x\right)^2_{n_c, n_d} = \langle (J_x)^2\rangle_{n_c, n_d} - \langle J_x\rangle^2_{n_c, n_d},
\end{equation}
where $\langle (J_x)^2\rangle_{n_c, n_d} = \mathrm{Tr}\left[J^2_x\rho_{n_c,n_d}\right]$. 

%%%==========================================================
\section{Zero tunneling and zero decoherence limit \label{sec:simplemodel}}
%%%========================================================== 
\subsection{General wavefunction after measurement}
%%%========================================================== 
In order to demonstrate how squeezing can be achieved using measurement, we consider first the limit where the tunneling is negligible so that we set $\Omega = 0$ in (\ref{eq:strn11}). This limit corresponds to a separation between the well minima that is very large in comparison to characteristic length of the oscillator in each well. Coherent light is used to measure the BEC in the double-well trap as shown in Fig.~\ref{fig1}. The quantum state of laser light in the two arms of the interferometer is  a coherent state of the form $\lvert \alpha_l\rangle\otimes\lvert\alpha_r\rangle$, where 
\begin{equation}
\label{eq:sm01a}
\lvert \alpha_{s}\rangle = e^{-\frac{\lvert \alpha_{s}\rvert^2}{2}}  e^{\alpha_s \hat{a}^\dagger_s}\lvert 0\rangle,
\end{equation}
$s=r,l$, $\lvert 0\rangle$ is the vacuum state. A generalized state of the BEC is the $J_x$ Fock basis $\lvert k, N-k\rangle$, and is written as
\begin{equation}
\label{eq:sm02a}
\lvert \Psi_0 \rangle = \sum_k C_k \lvert k , N- k\rangle,
\end{equation}
where $\lvert k , N- k\rangle$ represents the state having $k$ atoms in the left well and $N-k$ atoms in the right well,
\begin{equation}
	\label{eq:strn:me11}
	\lvert k , N- k\rangle = \frac{(b_l^\dagger)^k}{\sqrt{k!}}\frac{(b_r^\dagger)^{N-k}}{\sqrt{(N-k)!}}\lvert0\rangle\otimes\lvert0\rangle.
\end{equation}

The evolution of the atom-light state is governed by the Schr\"odinger equation $\lvert \Psi(t)\rangle = e^{-iH_\mathrm{eff}t/\hbar} \lvert\alpha_l\rangle\otimes\lvert\alpha_r\rangle\otimes\lvert\Psi_0\rangle$ and which can be solved exactly as 
\begin{align}
\label{eq:sm01}
\lvert \Psi(t) \rangle & = \lvert\alpha_l e^{-igtJ_x}\rangle\lvert\alpha_r e^{igtJ_x}\rangle\lvert \Psi_0 \rangle,\nonumber\\
&= e^{-\frac{|\alpha_l|^2}{2}-\frac{|\alpha_r|^2}{2}} \sum_k C_k\exp\left(\alpha_le^{-igt(k -N/2)}\hat{a}^\dagger_l\right)\nonumber\\ &\times\exp\left(\alpha_re^{igt(k-N/2)}\hat{a}^\dagger_r\right)\lvert 0\rangle\otimes\lvert 0\rangle\otimes\lvert k,N-k\rangle,
\end{align} 
where in the second equality we have expanded the BEC state in Fock basis~(\ref{eq:sm02a}). The phase of the photons encodes the information of the atoms. This phase information carried by the photons can be accessed via interference of the photons by passing the light through beamsplitter as shown in Fig.~\ref{fig1}, which is equivalent to the following transformation
\begin{equation}
\label{eq:strn:me07}
\hat{a}^\dagger_l = \frac{\hat{a}^\dagger_c + i\hat{a}^\dagger_d}{\sqrt{2}},\qquad \hat{a}^\dagger_r = \frac{i\hat{a}^\dagger_c + \hat{a}^\dagger_d}{\sqrt{2}}.
\end{equation}
The state after the beamsplitter becomes
\begin{align}
\label{eq:sm02}
\lvert\psi_\mathrm{BS}\rangle & = e^{-\frac{|\alpha_l|^2}{2}-\frac{|\alpha_r|^2}{2}}\sum_k C_k \exp\left(\frac{\alpha_c(k)}{\sqrt{2}} \hat{a}^\dagger_c\right)\nonumber\\
&\times \exp\left(\frac{\alpha_d(k)}{\sqrt{2}} \hat{a}^\dagger_d\right) \lvert0\rangle\otimes\lvert 0\rangle\otimes\lvert k,N-k\rangle,
\end{align}
where
\begin{align}
\label{eq:sm04}
\alpha_{c}(k) & =\left(\alpha_l + i\alpha_r\right) \cos[gt(k-N/2)] \nonumber\\
& - (i\alpha_l + \alpha_r)\sin[gt(k-N/2)],\\
\label{eq:sm05}
\alpha_{d}(k) & = (i\alpha_l +\alpha_r)\cos[gt(k-N/2)] \nonumber\\
& +(\alpha_l + i\alpha_r)\sin[gt(k-N/2)].
\end{align}

After the beamsplitter, the photons collected in the detectors $c$ and $d$ are counted. The probability $P(n_c,n_d)$ that $n_c$ and $n_d$ photons are detected at the detectors $c$ and $d$, respectively, is obtained by projecting  the operator 
$\lvert n_c, n_d\rangle\langle n_c , n_d\rvert $ on the state $\lvert\psi_\mathrm{BS}\rangle $,
\begin{align}
\label{eq:sm03}
P(n_c,n_d) & = \sum_{k} \lvert\langle n_c,n_d, k \lvert\psi_\mathrm{BS}\rangle\rvert^2,\nonumber\\
& =    \sum_{k} \lvert C_k\rvert^2 \lvert A_{n_c,n_d}(k)\rvert^2.
\end{align} 
The state $\lvert\psi_{n_c,n_d}\rangle$ after $n_c$ and $n_d$ photons have been detected is
\begin{equation}
\label{eq:sm06}
\lvert\psi_{n_c,n_d}\rangle = \frac{1}{\sqrt{P(n_c,n_d)}} \sum_{k} C_k A_{n_c,n_d}(k)\lvert k, N-k\rangle
\end{equation}
where 
\begin{equation}
\label{eq:sm06a}
A_{n_c,n_d}(k) = {e^{-\frac{|\alpha_l|^2}{2}-\frac{|\alpha_r|^2}{2}}} \frac{\left(\dfrac{\alpha_{c}(k)}{\sqrt{2}}\right)^{n_c}}{\sqrt{n_c!}}  \frac{\left(\dfrac{\alpha_{d}(k)}{\sqrt{2}}\right)^{n_d}}{\sqrt{n_d!}}.
\end{equation}
Thus the effect of measurement is to modify the initial probability amplitude of the BEC state, $C_k \rightarrow C_k A_{n_c,n_d}(k)$. 

In the limit $n_d,\,n_c\gg1$ and $gt >0$, the approximate state of the BEC after the photon detection becomes
\begin{equation}
\label{eq:sm07}
\lvert \psi_{n_c,n_d}^{\mathrm{approx}}\rangle \propto \sum_k C_k e^{-\frac{X_0}{4}(k- N/2 - x_0)^2}\lvert k, N-k\rangle,
\end{equation}
where the peak and width of the measurement weighting function $A_{n_c,n_d}(k)$ are, respectively 
\begin{align}
\label{eq:sm09}
x_0 &= \frac{1}{2gt}\left(\phi - \arcsin\left[ \frac{|\alpha_l|^2 + |\alpha_r|^2}{2|\alpha_l\alpha_r|}\frac{n_c - n_d}{n_d + n_c}\right]\right),\\
\label{eq:sm10}
X_0& = (gt)^2\left(\frac{n_c + n_d}{n_cn_d}\right)\Bigg[(n_c + n_d)^2\left(\frac{2|\alpha_l\alpha_r|}{|\alpha_l|^2 + |\alpha_r|^2}\right)^2\nonumber\\ 
&- (n_d - n_c)^2 \Bigg].
\end{align}
Detailed derivations of these equations above state are given in Appendix \ref{sec:approximatestate}.  From (\ref{eq:sm07}), we see that the factor $\lvert A_{n_c,n_d}(k)\rvert$ (\ref{eq:sm06a}) modifies the original probability amplitude $C_k$ by suppressing the amplitudes away from $ k = N/2-x_0 $. 

This clearly shows that the measurement process induce squeezing effect for a wide range of initial states. To illustrate the effect of measurement $A_{n_c,n_d}$, let us examine what happens to the probability amplitude of the BEC being in the state $\lvert k, N-k\rangle$ after measurement of $n_c$ and $n_d$ photons. In the limit of no atom-light interaction with $gt =0$, $A_{n_c,n_d}(k)$ is independent of $k$, and the initial state of the BEC is recovered. For a spin coherent state, the probability amplitude $C_k$ is characterized by a distribution that is Gaussian with a width scaling as $\sqrt{N}$. After the measurement with finite $gt$, the probability amplitude is modified as $C_kA_{n_c,n_d}(k)$. The state is squeezed if this new probability amplitude $C_kA_{n_c,n_d}(k)$ has a width that is smaller than $\sqrt{N}$.  

%=======================================================
\subsection{Example: initial spin coherent state \label{sec:sec:example}} 
%=======================================================
We now consider the specific case where the initial state of the atoms are spin-polarized in the $J_x$-direction. The full calculations are shown in Appendix C.  The final approximate state for $ n_c, n_d \gg 1 $ is 
\begin{align}
\label{eq:sma08}
\lvert \psi^\mathrm{approx}_{n_c,n_d}\rangle \approx& \left(\frac{1}{2\pi \sigma}\right)^{1/4}\nonumber \sum_k e^{ - \frac{1}{4\sigma}
	\left(k -   k_0\right)^2}\\
\times& e^{ik\varphi}e^{i(n_c\phi_c(k) + n_d\phi_d(k))}\lvert k, N-k\rangle,
\end{align}
where $\varphi$ is the initial relative phase between the atoms in the left and right well, $\phi_c(k)$ is the argument of $\alpha_c(k)$, and $\phi_d(k)$ is the argument of $\alpha_d(k)$. The general dependence of the width $\sigma$ and peak $k_0$ of the modulating Gaussian in (\ref{eq:sma08}) on detected number of photons $n_c$ and $n_d$ are derived in Appendix~\ref{sec:photonprobability}. For the case where $n_d = n_c$, the width $\sigma$ and the peak $k_0$ of the magnitude of  (\ref{eq:sma08}) are
\begin{align}
\label{eq:sma11}
\sigma = & \frac{N\lvert\eta_l\eta_r\rvert^2}{1 + 8g^2t^2n_cN\lvert\eta_l\eta_r\rvert^2},\\
\label{eq:sma12}
k_0 =& \frac{N|\eta_l|^2 + 4g^2t^2n_cN^2|\eta_l\eta_r|^2}{1 + 8g^2t^2n_cN|\eta_l\eta_r|^2 },
\end{align}
respectively, where   $\eta_l = (\alpha + \beta)/\sqrt{2}$, and  $\eta_r = (\beta - \alpha)/\sqrt{2}$. It immediately follows from (\ref{eq:sma08})  that the conditional probability density becomes 
\begin{align}
\label{eq:sma09}
P(k\lvert n_c) & = \sqrt{\frac{1}{2\pi \sigma}} e^{-\frac{1  }{2\sigma} \left(k -   k_0\right)^2}. 
\end{align}
Clearly, the state (\ref{eq:sma08}) is squeezed since for $gt\neq 0$, the width  of the exponential term $\Delta k =\sqrt{\sigma} $ is smaller than $\sqrt{N} \lvert\eta_l\eta_r\rvert$, the width (standard deviation) of the spin coherent state.
%
%%===========================================
%%Figure 2
%%===========================================
\begin{figure}[t]
	\includegraphics[width=\columnwidth]{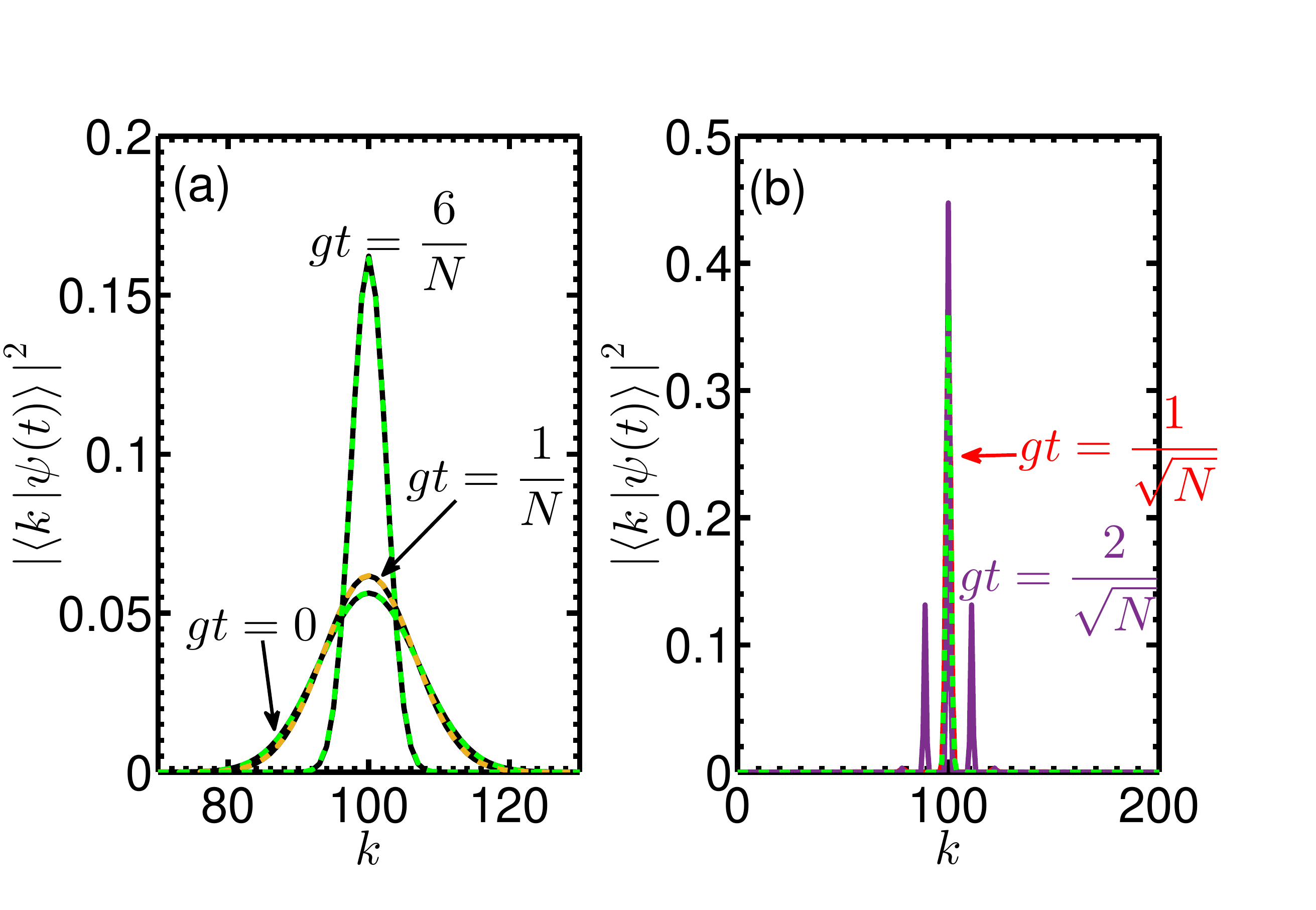}
	\caption{The conditional probability of finding the atoms in the state $\lvert k\rangle$. (a) The solid line is calculated numerically from (\ref{eq:sm06}) while the dashed line is given by (\ref{eq:sma09}). (b) The numerically computed probability density from (\ref{eq:sm06}) at large values of $gt$, $gt \sim 1/\sqrt{N}$ and the dashed lines is given by (\ref{eq:sma09}). The parameters for the plots are $N = 200$, $\alpha_l = \alpha_r = \sqrt{20}$, $ \alpha = 0$, and $\beta = 1$. }
	\label{fig2}
\end{figure}
%%===========================================
%

To illustrate this, (\ref{eq:sma09}) is plotted in Fig~\ref{fig2}. It shows excellent agreement with the exact calculation based on (\ref{eq:sm06}). In the presence of interaction, (\ref{eq:sma09}) shows that the peak of the BEC's probability distribution given that $n_c$ and $n_d$ photons have been detected is $k_0 $. For $gt = 0$, the BEC's probability distribution is that of a coherent state that has a peak at $N\lvert\eta_l\rvert^2$ and a width $\Delta k = \sqrt{N} \lvert\eta_l\eta_r\rvert$. With atom-light interactions $gt\neq 0$, the width of the probability density is modified as $\Delta k = \sqrt{\sigma} $. Clearly this shows that the width of the distribution decreases since the denominator of (\ref{eq:sma11}) would always be greater than unity if $gt$ is not zero. In the small interaction regime shown in Fig~\ref{fig2}(a), $gt \ll 1/\sqrt{N}$, the width decreases from that of a coherent state by an amount $4g^2t^2\lvert\alpha_r\rvert^2N^{3/2}|\eta_l\eta_r|^3$. Since the width of the BEC probability density distribution in a conditional measurement of $n_c$ and $n_d$ photons is smaller than $\sqrt{N}|\eta_l\eta_r|$, the BEC state is thus squeezed. For $gt \sim 1/\sqrt{N}$ the approximate probability (\ref{eq:sma09}) does not predict the many peaks that emerge in this limit. Thus the approximate probability breaks down and is no longer valid. The numerically computed results predict that the squeezing is lost as the width of the probability density distribution is roughly $\Delta k \sim N$. This is clearly supported by the results of Fig.~\ref{fig2}(b) which shows there is more than one peak in this limit which results in the broadening of the width of the distribution. We remark that the inequality $gt \ll 1/\sqrt{N}$ is an upper bound for the atom-light interaction strength beyond which squeezing is being lost. It is possible to obtain a much lower bound on atom-light interaction strength that would depend on the average photon number used in the experiment.

\begin{figure}[t]
	\includegraphics[width=\columnwidth]{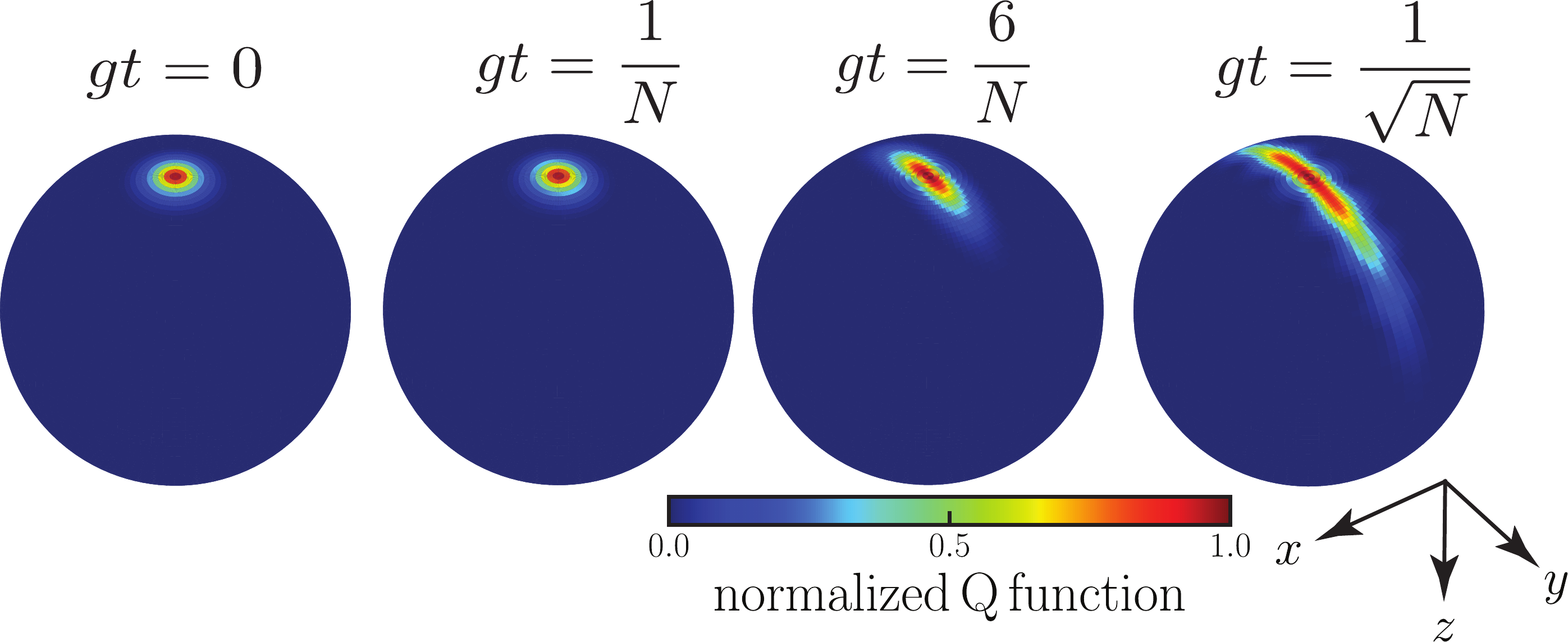}
	\caption{The conditional Husimi \emph{Q} function. The parameters for the plots are $N = 200$, $\lvert\alpha_l\lvert = \lvert\alpha_r\rvert = \sqrt{20}$, $ \alpha = 0$, and $\beta = 1$. The parameter $gt$ is as shown in the figure. For ease of visualization, the positive \emph{z}-axis is flipped and points downwards.}
	\label{fig3}
\end{figure}

%=========================================================
%\subsection{Husimi \emph{Q}-Distribution \label{sec:sec:simpleHusimi}}
%=========================================================
The BEC state after measurement can also be visualized on a Bloch sphere using the Husimi \emph{Q} function
\begin{equation}
\label{eq:sm11}
Q = \frac{N+1}{4\pi} \lvert\langle\theta,\phi\rvert\psi_{n_c,n_d}\rangle\lvert^2,
\end{equation}
where $\lvert\theta,\phi\rangle$ is the atomic coherent state (\ref{eq:strn:me06}). The results of plotting (\ref{eq:sm11}) in Fig.~\ref{fig3} show that for $gt = 0$, one obtains a distribution that has equal width in $\theta$ and $\phi$, which is characteristic of a spin coherent state. As the atom-light interaction strength $gt$ increases, $gt\sim 1/N$, the \emph{Q}-function distribution takes the shape of an ellipse that is squeezed along \emph{x}-axis as shown at $gt =6/N$. For strong atom-light interaction strength $gt\sim 1/\sqrt{N}$, the squeezing starts to degrade as the ellipsoid shape is destroyed. There appears broader distribution around the ellipse that distorts its shape, and the distribution starts to wrap around the Bloch sphere. This can be linked to the many connected peaks that emerge in this limit as shown in Fig~\ref{fig2}(b).  As result, the width of the distribution broadens and is greater than $\sqrt{N}$. 

%%===========================================
%%Figure 3
%%===========================================
%%===========================================

These calculations show that measurement can be used to create spin squeezed states in an atomic BEC that is trapped in a double-well potential. Also, we accounted for how the squeezing is lost due to strong measurement. In the following section we model the system for the more general case, accounting for the effects of dephasing and tunneling.

%%%==========================================
\section{Tunneling with zero decoherence\label{sec:blochspheretunneling}}
%%%==========================================
In this section we calculate squeezing in the BEC state for measurements performed on the condensates while tunneling between the wells is present  but without any dephasing. The resulting effect of the interactions on quantum state of the BEC is then analyzed by studying its evolution on the Bloch sphere. Similarly, the effects of measurement on the BEC state are analyzed via the evolution of spin operators.

In the presence of tunneling alone without measurement, the initial state (\ref{eq:strn07}) evolves  within the mean field description as $ \lvert\psi(t) \rangle = e^{-i\hat{H}t/\hbar}\lvert\alpha,\beta\rangle$. Defining the relative phase of the atom probability amplitudes $\phi_{\alpha,\beta} =\phi_{\alpha} - \phi_{\beta}$, then the expectation values of the spin operators take a simple form
\begin{align}
	\label{eq:strn09}
	\langle\psi(t)\lvert J_x\rvert\psi(t)\rangle & = N\lvert\alpha\beta\rvert\cos(\Omega t - \phi_{\alpha,\beta}),\nonumber\\ \langle\psi(t)\lvert J_y\rvert\psi(t)\rangle & = N\lvert\alpha\beta\rvert \sin(\Omega t - \phi_{\alpha,\beta}),\\ 
	\langle\psi(t)\rvert J_z\lvert\psi(t)\rangle & = N\frac{\lvert\alpha\rvert^2 - \lvert \beta\rvert^2}{2}.\nonumber
\end{align}
The corresponding variances of the atomic spin operators are readily calculated in a similar way 
\begin{align}
	\label{eq:strn10}
	(\Delta J_x)^2 = & \frac{N}{4}\left[1 - 4\lvert\alpha\beta\rvert^2\cos^2(\Omega t -\phi_{\alpha,\beta})\right]\nonumber,\\
	(\Delta J_y)^2 = & \frac{N}{4}\left[1 - 4\lvert\alpha\beta\rvert^2\sin^2(\Omega t -\phi_{\alpha,\beta})\right],\\
	(\Delta J_z)^2 = & N\lvert\alpha\beta\rvert^2.\nonumber
\end{align}
The tunneling causes the atoms to oscillate between the two wells which on the Bloch sphere is represented by a precession about the \emph{z}-axis with frequency $\Omega$ as expressed in (\ref{eq:strn09}). Since the squeezing occurs in the $ J_x $-direction, this means that the tunneling causes the state to rotate with respect to the squeezing direction. As such, one strategy is to perform measurement at specific times $t=n\pi/\Omega$, where the oscillator strength is maximum, in order to generate squeezing~\cite{thorne1978,Braginsky1980,vasilakis2015}. Alternatively, one could perform continuous measurement on the oscillator and generate a squeezed state, for sufficiently strong enough atom-light coupling strength, using the balanced detection scheme described in Sec.~\ref{sec:sec:example}.  The latter approach is what we study in this section.

%%%==========================================
\subsection{Husimi \emph{Q}-Distribution\label{sec:sec:tunnelingnodecoherence}}
%%%==========================================
The effect of atom-light interactions on the BEC state in the presence of tunneling without decoherence is examined by numerically  calculating the density matrix $\rho$ using (\ref{eq:strn:me01}), with the initial state $\rho_0 =\lvert\alpha,\beta\rangle\langle \alpha,\beta\rvert$, where $\lvert\alpha,\beta\rangle $ is given in (\ref{eq:strn07}). At the end of evolution, the photons are recombined using (\ref{eq:strn:me07})  and subsequently detected at the detectors \emph{c} and \emph{d}. The state $\rho_{n_c,n_d}$ (\ref{eq:strn:me10}) resulting from the detection of $n_c$ and $n_d$ photons is then used to calculate the Husimi \emph{Q}-function defined for a mixed state as 
\begin{equation}
\label{eq:papc05}
Q_{n_c,n_d}(\theta,\phi,t) =  \frac{N+1}{4\pi}\langle\theta,\phi\lvert\rho_{n_c,n_d}(t)\lvert\theta,\phi\rangle,
\end{equation}
where $\lvert\theta,\phi\rangle$ is a spin coherent state given in (\ref{eq:strn:me06}). %(\ref{eq:papc06}).

\begin{figure}[t]
	\includegraphics[width=\columnwidth]{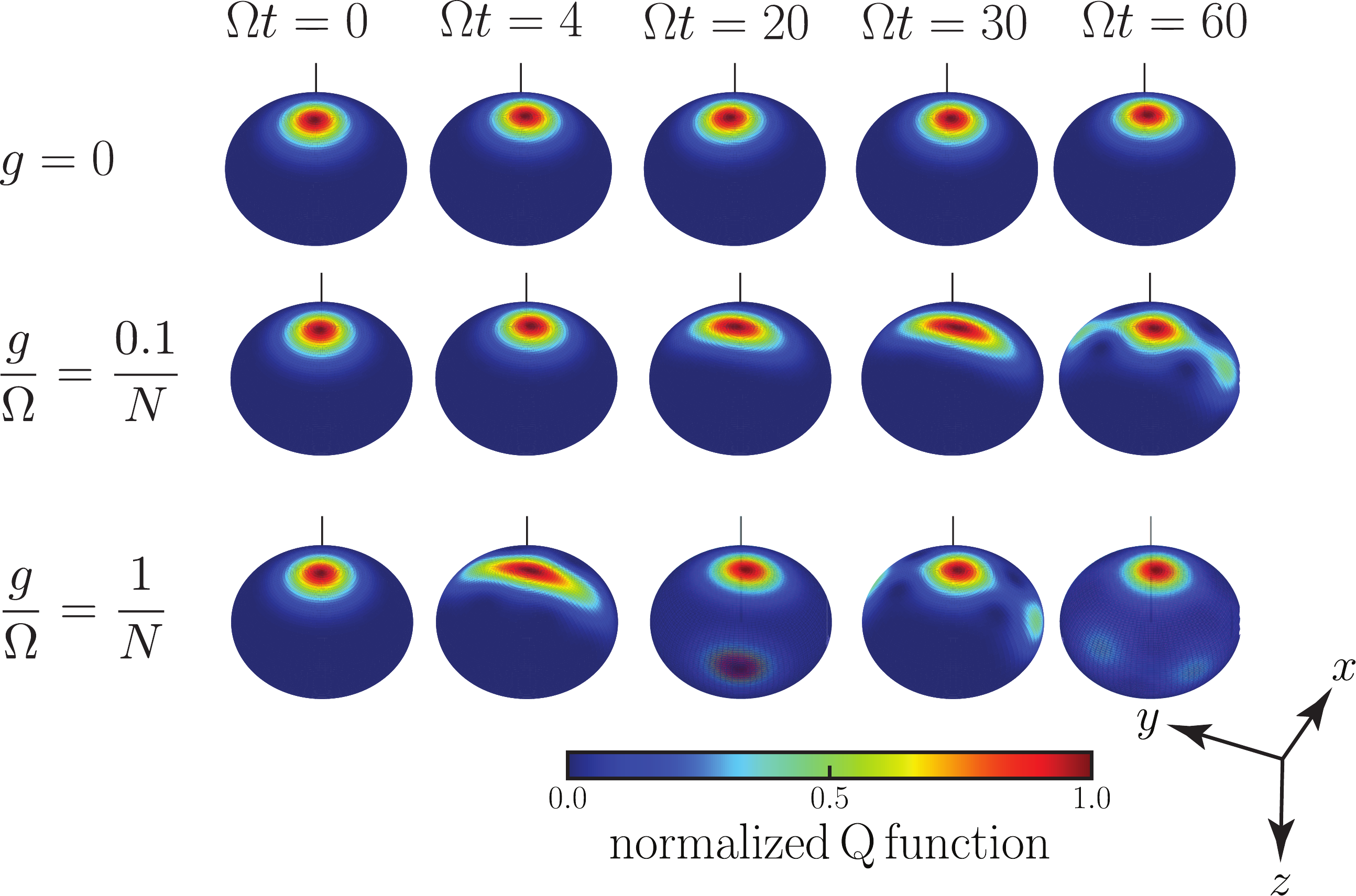}
	\caption{The conditional Husimi \emph{Q}-function. The parameters for the plots are $N = 30$, $\alpha = \sqrt{0.001}$, $\beta = \sqrt{0.999}$, $\Omega = \pi/4$ and $\alpha_l = \alpha_r = 2$. The parameters for the strength of the coupling $g$ and time $\Omega t$ are as shown in the figure. For ease of visualization, the positive \emph{z}-axis is flipped and points downwards.}
	\label{fig4}
\end{figure}

The results are shown in Fig.~\ref{fig4}. The rows give the variations of the \emph{Q}-functions with time $\Omega t$, while the columns give the variation of the state $\rho_{n_c,n_d}$ with atom-light interactions $g/\Omega$. The first row at $g=0$ confirms that state $\rho_{n_c,n_d}$ is that of a spin coherent state that is precessing  about the \emph{z}-axis on a Bloch sphere. Due to the slight offset from a perfectly \emph{z}-polarized state, the  tunneling causes coherent state to rotate around the \emph{z}-axis. Also we see from the figure that weak atom-light interactions give rise to a slow build up of squeezing of the BEC state. For instance at $\Omega t = 4$, the BEC state is still similar to that of a coherent state at $g=0.1\Omega/N$ compared to that at $g = \Omega/N$ where the squeezing is already starting to degrade. At long times, the squeezing of the BEC state arising from a weak atom-light interactions is lost as shown for $\Omega t = 60$ and $g = 0.1\Omega/N$. This loss occurs because of the squeezed spin state transitioning into other quantum state as the interactions become stronger. We are able to probe this long interaction time region because we use the exact wavefunction approach. This is best seen in the case of strong atom-light interactions $g = \Omega/N$ where the loss of squeezed state after $\Omega t = 4$ leads to the emergence of two spin coherent state at the north and south poles of the Bloch sphere for  $\Omega t = 20$. Such a state is a Schr\"odinger cat state, which is a linear superposition of two spin coherent state~\cite{dodonov1974,byrnes2020}. Beyond the time $\Omega t = 20$, the spin coherent state at the top of the Bloch sphere splits into two and starts its motion back to the bottom of the sphere where they again form a spin squeezed state and then a spin coherent state (see also movies in the Supplementary Materials).   

%%===========================================
%%Figure 4
%%===========================================
%%=========================================== 

%%%==========================================
\subsection{Expectation values and variances\label{sec:sec:averagestunneling}}
\begin{figure}[t]
	\includegraphics[width=\columnwidth]{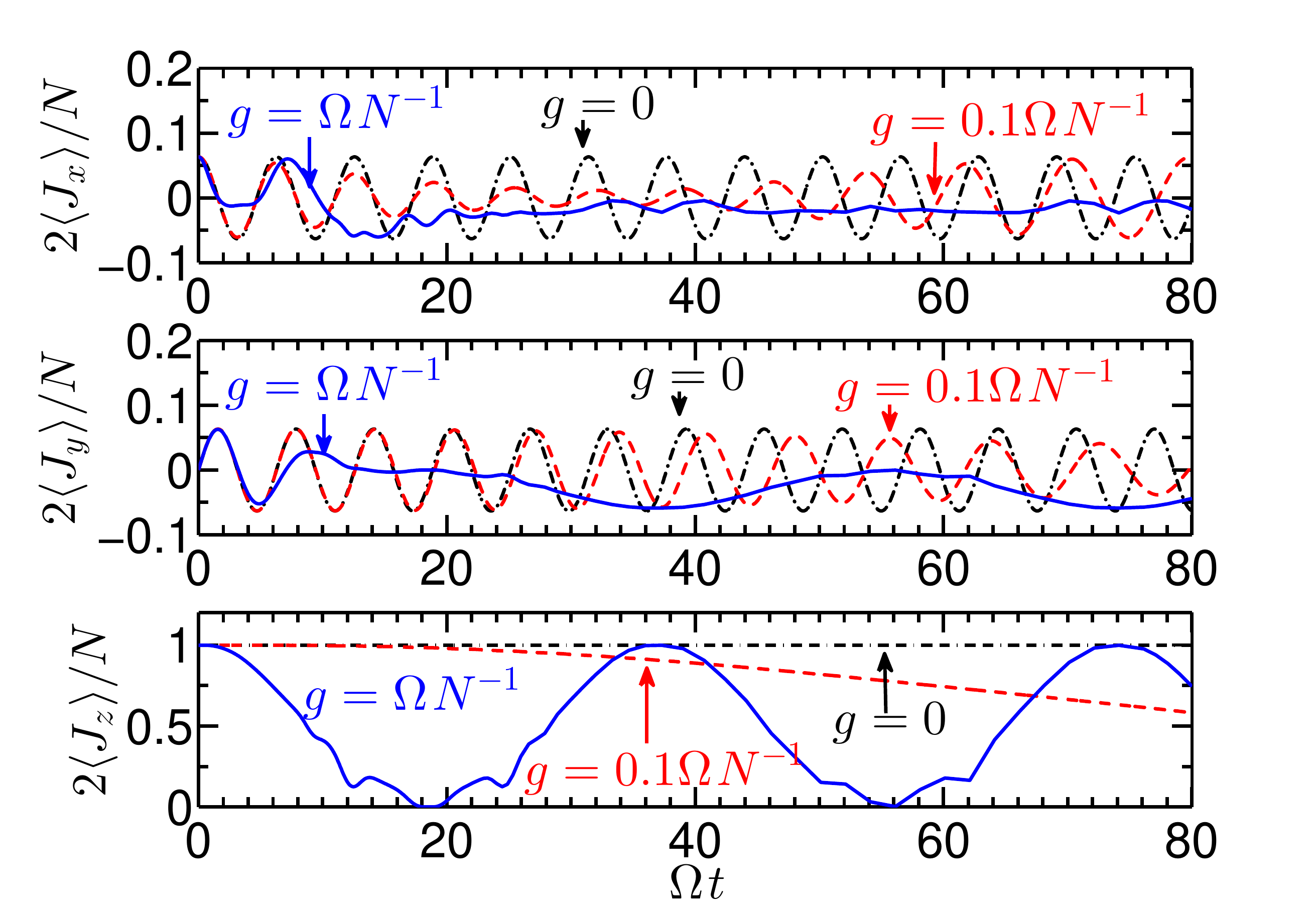}
	\caption{The normalized conditional mean of $J_x$, $J_y$ and $J_z$ with dephasing parameter $\gamma$ = 0. The atom-light interaction strengths are as shown in the figure. The parameters for the figure are $N =30$, $\Omega = \pi/4$, $\alpha =\sqrt{0.001}$, $\beta=\sqrt{0.999}$, $\alpha_{r}=\alpha_{l}=2$.}
	\label{fig5}
\end{figure}

\begin{figure}[t]
	\includegraphics[width=\columnwidth]{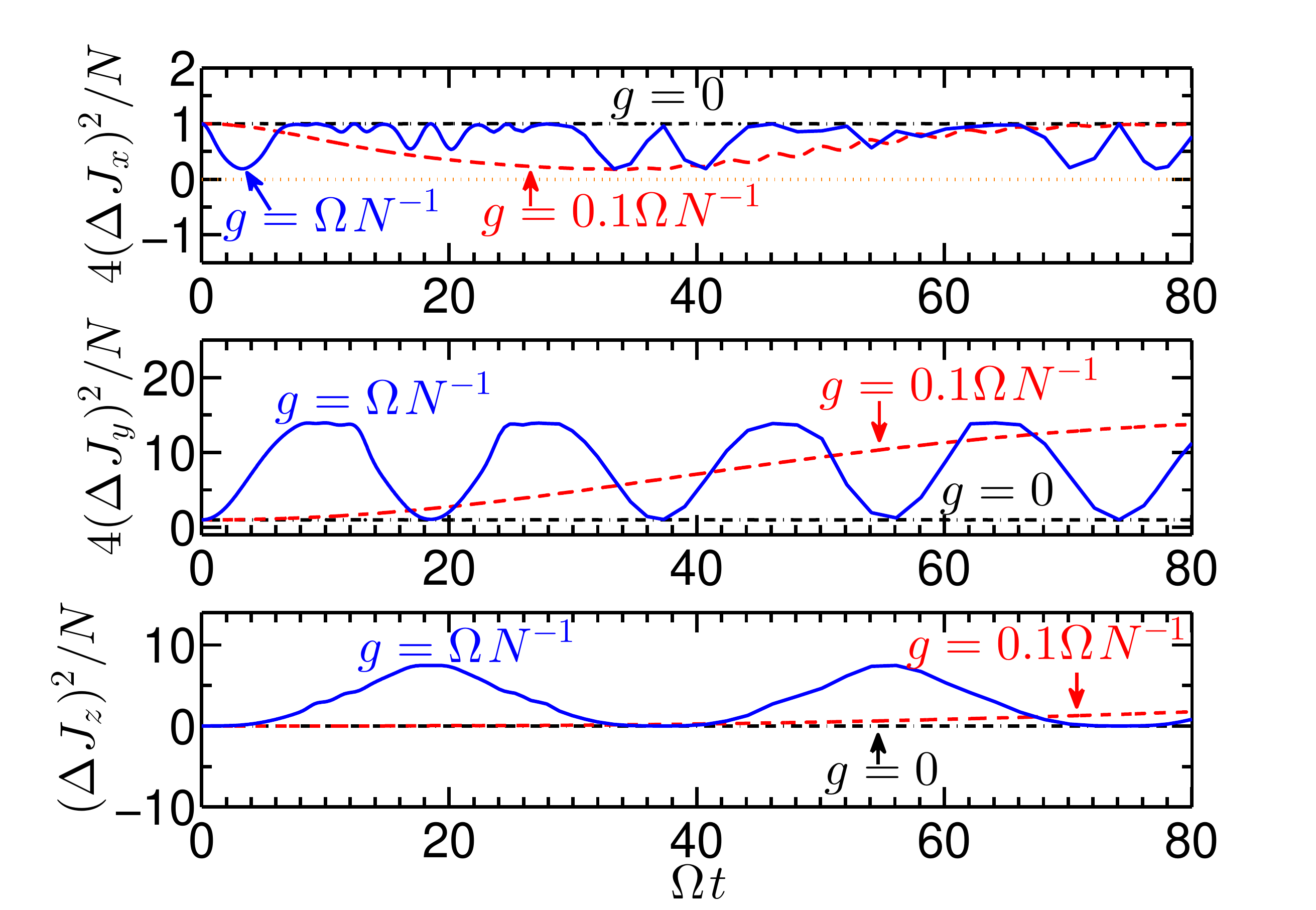}
	\caption{The normalized conditional variance of $J_x$, $J_y$ and $J_z$ at dephasing parameter $\gamma$ = 0. The atom-light interaction strengths are as shown in the figure. The dotted line at normalized conditional variance of $J_x$ equal to zero is a guide for the eye. The parameters for the figure are $N =30$, $\Omega = \pi/4$, $\alpha =\sqrt{0.001}$, $\beta=\sqrt{0.999}$, $\alpha_{r}=\alpha_{l}=2$.}
	\label{fig6}
\end{figure}

%%%==========================================
We now examine the spin averages and variances with the tunneling present. The averages calculated at different atom-light coupling strength $g$ with the dephasing parameter  $\gamma$ set to zero is shown in  Fig.~\ref{fig5} and Fig.~\ref{fig6}. For $g = 0$,  the expectations  $J_x$  and $J_y$ execute harmonic oscillation with a relative phase of $\pi/2$  between them. As a result, the mean spin projection on the \emph{x}-\emph{y} plane traces a circle thereby confirming that  the atoms precess about the \emph{z}-axis on the Bloch sphere at frequency $\Omega $. Physically this corresponds to the atoms shuttling between the two double-well trap at frequency $\Omega$. Also, the relative atomic population in the excited state remains fixed for all times. Since most of the atoms are initialized to be dominantly in the ground state of the double-well, the variance of $J_z$ remains fixed as predicted by (\ref{eq:strn09}), as shown in Fig.~\ref{fig5}. However, the variances of $J_x$ and $J_y$ are large, i.e. around unity in scaled units, see (\ref{eq:strn10}).  
%%===========================================
%%Figure 5
%%===========================================
%%===========================================

Turning on the measurement apparatus with coupling strength $g = 0.1\Omega/N$ in Fig.~\ref{fig5}, the atoms execute sinusoidal oscillations between the wells with a slightly modulated frequency.  This can be understood from the fact that the first order effect is to change the Hamiltonian to $H = \hbar\Omega J_z + \hbar g \langle S_x \rangle J_x$ which changes the axis of the oscillations. The amplitude of $\langle J_x\rangle$ oscillations decreases faster than the amplitude of $\langle J_y\rangle$ oscillations  as shown by the dashed lines in Fig.~\ref{fig5}.  As a result,  the circular motion traced by mean spin projection on the \emph{x}-\emph{y} plane shrinks faster along \emph{x} than along \emph{y} giving rise to an ellipsoidal trajectory. This squeezing effect is manifested in the variance of $J_x$ which decreases below unity, while that of $J_y$ increases well above unity as shown by the dashed line in Fig.~\ref{fig6}. The variance of $J_x$ reaches its minimum value in the region where the amplitude of $\langle J_x\rangle$ oscillations has its smallest value. Observing the variance to be below unity indicates that the state of the oscillator is squeezed. At the same time there is oscillation in both the \emph{y}-\emph{z} and \emph{x}-\emph{z} planes which is absent in the case without measurement $g=0$. We also point out that there are two time scales. One of them is the fast oscillation responsible for the tunneling between the wells. In addition, there is modulation of fast oscillations that is responsible for the squeezing effect. Comparing the frequencies in the expectations of $J_x$ and $J_y$ at $g=0$ and $g = 0.1\Omega/N$, it is seen that atom-light interactions increases the period of the fast time scale for $\Omega t >20$.   

Increasing further the atom-light interaction strength to $g = \Omega/N$ as represented by the dotted lines in Fig.~\ref{fig5} shows that the motion is no longer periodic after a short time. However,  the variance of $J_x$ is again well below unity, while that of $J_y$ is well above unity within the short time window as shown in Fig.~\ref{fig6}. Notice that the squeezing of the state is more easily seen with the variance of $J_x$ than in the \emph{Q}-function. Note that the minimal width of a $Q$-function, even for an infinitely squeezed state, is only one-half that of a spin coherent state.
%%===========================================
%%Figure 6
%%===========================================
%%===========================================

%%%==========================================
\section{Tunneling and dephasing effects \label{sec:blochspheredephasing}}
%Generalized Bloch sphere: Effect of ac Stark shift dephasing
%%%==========================================
We now include the dephasing, and investigate how the state of the BEC is impacted by visually inspecting changes in the plots of the \emph{Q}-function and more quantitatively by analyzing the averages of the spin operators. 

%%%==========================================
\subsection{Husimi \emph{Q}-Distribution\label{sec:sec:tunneldephasing}}
%%%==========================================
We begin by calculating the Husimi $Q$-function given in (\ref{eq:papc05}). The density matrix $\rho_{n_c,n_d}$ is given in (\ref{eq:strn:me10})  with $\Omega \geq0$ and $\gamma\geq0$. The results of the calculations are shown in Fig.~\ref{fig7}. Moving along the rows show variation of the \emph{Q}-functions for the state $\rho_{n_c,n_d}(t)$ with time, while moving along the column shows the variation  with dephasing rate $\gamma$. As time increases, there is significant squeezing of state $\rho_{n_c,n_d}(t)$. The state with weak dephasing rate shows a slow build up to a noticeable amount of squeezing. At longer times $\Omega t \geq 50$, the state can however lose its squeezing depending on the dephasing rate $\gamma$. The onset for the loss of squeezing happens earlier if the dephasing rate is very strong. For instance, for $\gamma = 3g$ the state loses its squeezing at $\Omega t = 30$ while for $ \gamma = g $ the squeezing is still in tact at the same time.  In the case of dephasing rate $\gamma = 0.1g $, the state exhibits strong squeezing at $\Omega t = 30$ which will begin to diminish at $\Omega t = 50$.

%%%==========================================
\subsection{Expectation values and variances }
%%%==========================================
%%===========================================
%%Figure 7
%%===========================================
\begin{figure}[t]
	\includegraphics[width=\columnwidth]{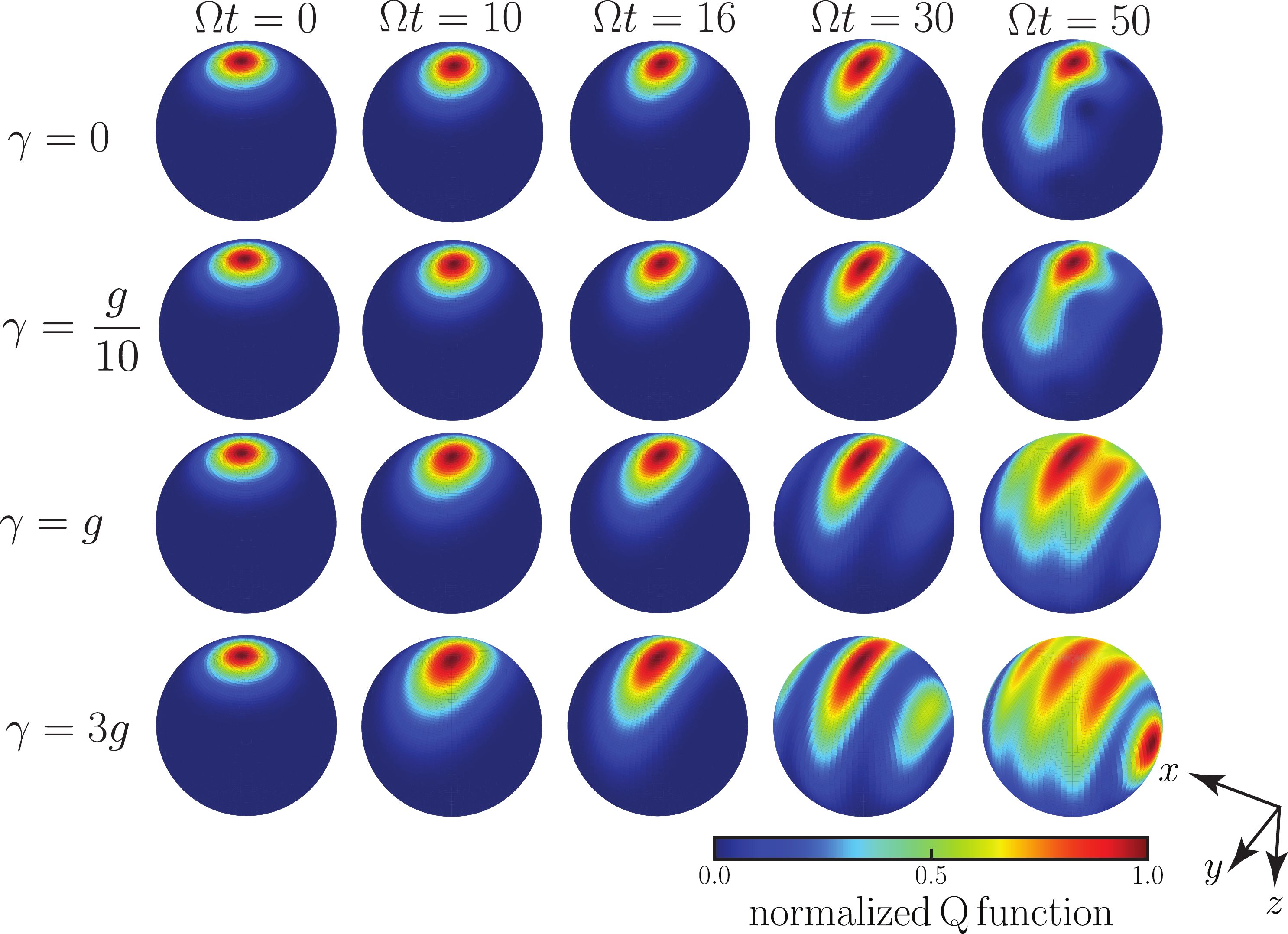}
	\caption{The conditional Husimi \emph{Q}-function. The parameters are $N=30$, $\Omega = \pi/4$, $g = 0.1\Omega/N$, $\alpha= \sqrt{0.001}$, $\beta = \sqrt{0.999}$, and $\alpha_l=\alpha_r =2$. The dephasing rate $\gamma$ (row) and the parameter proportional to time $\Omega t$ (columns) are as indicated. For ease of visualization, the positive \emph{z}-axis is flipped and points downwards. }
	\label{fig7}
\end{figure}
\begin{figure}[t]
	\includegraphics[width=\columnwidth]{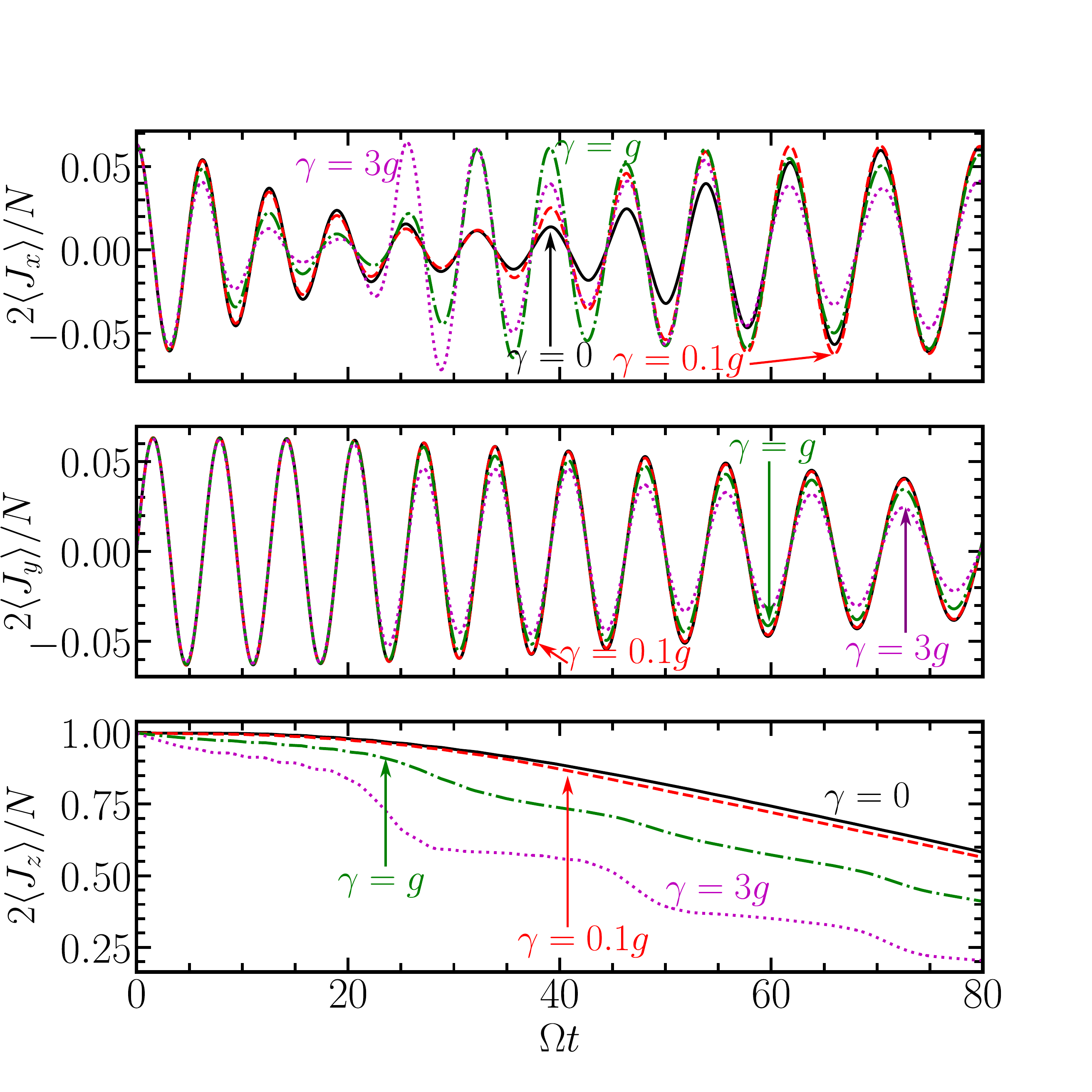}
	\caption{The normalized conditional mean of $J_x$, $J_y$ and $J_z$ at different values of dephasing parameter $\gamma$. The measurement strength $g = 0.1\Omega/N$ for all plots. The solid line is for $\gamma = 0$, the dashed line is for $\gamma= 0.1g$, the dashed-dotted line is for $\gamma = g$ and the dotted line is for $\gamma = 3g$. The parameters for the figure are $N =30$, $\Omega = \pi/4$, $\alpha =\sqrt{0.001}$, $\beta=\sqrt{0.999}$, $\alpha_{r}=\alpha_{l}=2$.}
	\label{fig8}
\end{figure}
\begin{figure}[t]
	\includegraphics[width=\columnwidth]{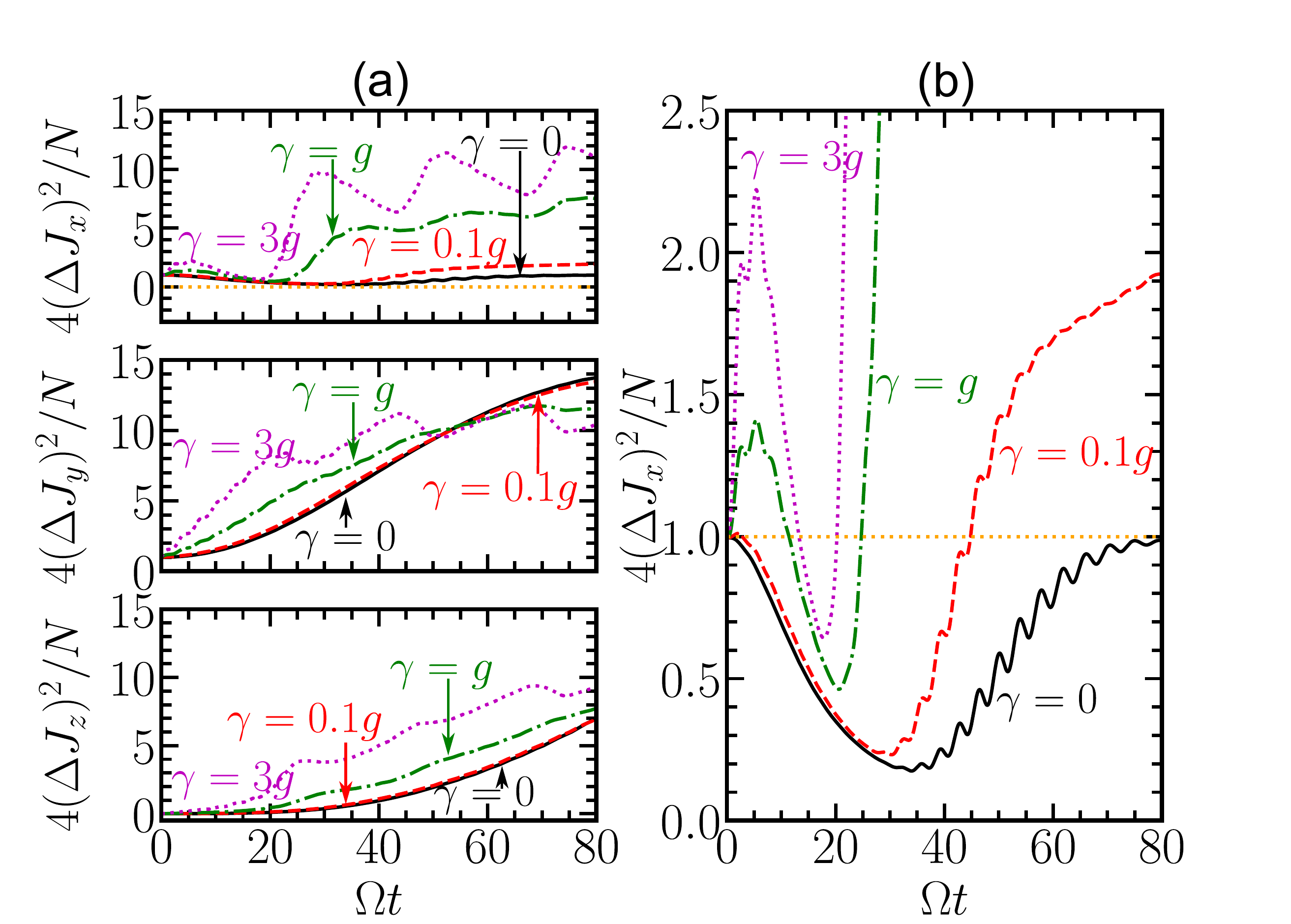}
	\caption{The normalized conditional variance of $J_x$, $J_y$ and $J_z$ at different values of dephasing parameter $\gamma$. (a) Presented in this column are the normalized variances of  $J_x$, $J_y$ and $J_z$. The dotted line at $4(\Delta J_x)^2/N =0$ is a guide for the eye. (b) An enlarged plot of the normalized variance of $J_x$  showing its behavior around unity for different values of dephasing parameter. The solid line at $4(\Delta J_x)^2/N =1$ is guide for the eye.   The parameters for the figures are $N =30$, $\Omega = \pi/4$, $g = 0.1\Omega/N$, $\alpha =\sqrt{0.001}$, $\beta=\sqrt{0.999}$, and $\alpha_{r}=\alpha_{l}=2$.} 
	\label{fig9}
\end{figure}

%%===========================================
Fig.~\ref{fig8} and Fig.~\ref{fig9} show the spin expectation values and variances for the case including dephasing, $\gamma \geq 0$. Additionally, Fig.~\ref{fig9}(b) contains an enlargement of the variance of $J_x$ around unity.  One effect of dephasing is to maintain the amplitude of the oscillations in $J_y$ while causing a decrease in the oscillation amplitude of the $J_x$ expectation value. Hence as the amplitude of $\langle J_x\rangle$ oscillations decrease and go through its minimum point, the amplitude of $\langle J_y\rangle$ oscillations is roughly same as that at $\gamma = 0$. Consequently, the mean spin projection on the $x$-$y$ plane does not trace an ellipse until after the minima of the $J_x$ amplitude has been passed and begins to increase. As such there is a delay in the decrease of the variance of $J_x$ below unity in the presence of dephasing compared to that when dephasing is absent $\gamma =0$ as shown in Fig.~\ref{fig9}(b). This effect becomes more noticeable for large dephasing rate $\gamma \geq g$. As a result,  the squeezing can persist in the presence of dephasing but only within a certain time window. Similarly, the minimum of the variance of $J_x$ is reached quite early for appreciable dephasing rate. Hence, a strong squeezing effect can occur quite early when the dephasing rate is appreciable while it happens much later if the dephasing rate is weak. These results are in agreement with that already shown in Fig.~\ref{fig7}. Also, note that the period of the fast oscillations is same for the different strength of the dephasing value i.e dephasing does not affect the effective tunneling period of the atoms.

%%===========================================
%%Figure 8
%%===========================================
%%=========================================== 

%%%========================================================== 
\section{Experimental realizations\label{sec:experiments}}
%%%========================================================== 

Nondestructive measurement of atoms have been carried out in several experiments~\cite{higbie2005,meppelink2010}. In these realizations, the measurement is performed on atoms in a single-well trap, where the single-site addressing is not required. Nondestructive measurement of atoms in double well requires independent access to each site. The double-well separation in cold-atom experiments is on the order of micrometers in order to have reasonable control of the quantum tunneling between the wells~\cite{schumm2005,albiez2005}. Diffraction limits the experimentally feasible waist size of the probe laser beam to be of same order corresponding to a few wavelengths. Hence it is difficult to obtain the arrangement shown in Fig.~\ref{fig1}.

However, for a  spinor BEC in the double-well trap, it is possible to separately access information about each site as described in~Ref.\cite{ilo-okeke2014}, where the effective Hamiltonian (\ref{eq:strn11}) can be recovered by adding a weak microwave coupling between internal states \cite{riedel2010}. In addition, the atoms in each well can be addressed separately by working in the frequency domain where the scattered light from atoms in the left and right wells are different to those of the incident light. This difference in frequencies, where, the atoms in the left well are engineered to modulate the incident light to lower frequency, while atoms in the right well absorb the incident beam and emit at higher frequency or vice versa, can then be used to distinguish the information from the atoms on the left well from those on the right well. Such a technique can be implemented using Raman transitions~\cite{kasevich1991,moler1992,carraz2012,giese2013} between the ground state energies of the atom, and the modulated output beam would be observed on a polarimeter.

%%===========================================
%%Figure 9
%%===========================================
%%===========================================

Alternatively, an extension of the RF-dressed Voigt detection method \cite{jammi2018} in multiple-RF dressed double-well potentials \cite{harte2018,barker2020} provides a convenient method of measurement explained in Sec. \ref{sec:sec:dephasing}. In this method, the double-well potential is formed by a combination of three RF dressing fields, and atoms at the two minima are on resonance  with maximum and minimum dressing RF components respectively. An off-resonant, polarized probe laser beam passing through both trapped clouds is phase modulated at the frequencies of the resonant dressing fields, and the signal at each frequency  can be detected with a balanced polarimeter \cite{jammi2018}. The detection of the two frequency components at the dressing field frequencies using two separate lock-in demodulations constitutes the measurement equivalent to that shown Fig.~\ref{fig1}.

%%%==========================================================
%%% Conclusion
%%%==========================================================
\section{Summary and Conclusions\label{sec:discussion}}
%%%========================================================== 
We showed that minimally-destructive detection is a good tool for the creation of a squeezed state of atoms in a double-well potential. To achieve the minimally-destructive effect on the states of the atoms, the measurement is performed with a balanced detection configuration. To understand the squeezing effect, we first studied the system under the conditions of no tunneling and no decoherence. Our main result in this case is (\ref{eq:sm06}), where a simple form of the atomic wavefunction after measurement was derived.  Here, it was found that the measurement introduces an additional factor of $A_{n_c, n_d}(k)$ into the atomic wavefunction, which can be approximated as a Gaussian. Without squeezing, the probability distribution of the quantum state of the atoms  is approximately a Gaussian with a width of roughly $\sqrt{N}$. Our scheme can affect a reduction in the width of the probability distribution of the quantum state of the atoms below unsqueezed value of $\sqrt{N}$ after conditional detection of photons (\ref{eq:sma11}). One of the key features of our approach is that we solve exactly the time evolution of the wavefunction under the QND interaction.  This contrasts to standard methods based on the Holstein-Primakoff approximation which are limited to short interaction times and an initial spin coherent state.  Thus, we have been able to investigate the full non-linear effect of the QND Hamiltonian, showing the generation of oversqueezed states and exotic states such as Schr\"odinger cat states.

In the presence of tunneling, the squeezing effect was seen as a decrease in the variance of the atom spin operator that was coupled to light. Additionally, there is a  modulation of the expectation values of the spin operators arising from measurement which is absent when there is no tunneling effect. We studied the effects of dephasing on the squeezed state due to the presence of light using the master equation.These calculations showed that dephasing has a negligible effect on squeezing when the atom-light interaction strength is greater than the dephasing rate. Even in the strong dephasing regime, where the dephasing strength is equal or greater than the measurement strength, we found that squeezing persists. 

We showed the overall effect of squeezing and dephasing on the quantum state of the atoms by visualizing it on a Bloch sphere using the Husimi \emph{Q}-distribution. In the case without tunneling and dephasing, the initial mean spin direction is not affected by the measurement. However, the measurement shrinks  the \emph{Q}-distribution by causing squeezing in the plane that is perpendicular to the mean spin direction, and  in particular shrinks only the width of the \emph{Q}-distribution along the spin axis that is coupled to light. In the presence of tunneling, the mean spin direction is relatively unaffected by the measurement for small atom-light interaction strength. There are also oscillations of the state about the mean spin direction, and for relatively strong atom-light interactions, the width of the \emph{Q}-distribution along the spin axis that is coupled to light decreases, which manifests in the squeezing effect. 

The squeezing effect observed in the presence of tunneling is a precursor to formation of different quantum states of the BEC such as the Schr\"odinger cat-state that is characterized by two \emph{Q}-distributions located on the opposite (north and south) poles on the Bloch sphere. Due to the oscillations, the state formation is actually reversible where the BEC state goes between the cat-sate and the spin coherent state, with the spin squeezed state being one of the intermediate states in the transition. This actually mimics the collapse and revival of quantum state that have been observed in other different oscillating systems like the Jaynes-Cummings model~\cite{gea-banacloche1990,buzek1992}, and the Josephson junction~\cite{milburn1997,kirchmair2013}. It is also analogous to the oscillatory effects observed~\cite{byrnes2013} in the two-axis squeezing of BEC where there is a periodic evolution between a coherent state and a maximal entangled state. The inclusion of the dephasing causes several other satellite \emph{Q}-distributions to emerge. The splintered \emph{Q}-distributions are along the spin axis that is coupled to light, a feature that is absent if the dephasing effect is zero. Additionally, there is a broadening of the width of the \emph{Q}-distributions along the spin axis coupled to light, linking the various \emph{Q}-distributions. Hence, while squeezing is present at small atom-light interactions in the presence of dephasing, all the features described previously about the \emph{Q}-function are not recovered, and squeezing is irreversibly lost in the strong dephasing regime or at large atom-light interactions in the presence of dephasing.

\acknowledgments
T. B. is supported by the Shanghai Research Challenge Fund; New York University Global Seed Grants for Collaborative Research; National Natural Science Foundation of China (Grant No. 61571301); the Thousand Talents Program for Distinguished Young Scholars (Grant No. D1210036A); and the NSFC Research Fund for International Young Scientists (Grant No. 11650110425); NYU-ECNU Institute of Physics at NYU Shanghai; and the Science and Technology Commission of Shanghai Municipality (Grant No. 17ZR1443600). E. O. I. O. acknowledges the  Talented Young Scientists Program (NGA-16-001) supported by the Ministry of Science and Technology of China. S. S. acknowledges Murata scholarship foundation, Ezoe foundation and Daishin foundation. The authors in Oxford carry out experimental work on cold atoms funded by EPSRC grant EP/S013105/1.

\appendix
%%==========================================================
\section{Solving The Master Equation\label{sec:masterequation}}
%%==========================================================
To solve for the evolution of the density matrix, we use the ansatz~\citep{garcia-ripoll2005,hussain2014} 
\begin{equation}
\label{eq:strn:me02}
\rho = \sum_{k,k'}\rho_{kk'}\lvert k;\alpha_{k,l},\alpha_{k,r}\rangle\langle k';\alpha_{k',l}, \alpha_{k',l}\rvert,
\end{equation}
where for each atom number state $\lvert k\rangle = \lvert k,N-k\rangle$ defined in the $J_x$ basis, there is a coherent state $\lvert\alpha_{k,s} \rangle= e^{\tfrac{-|\alpha_{k,s}|^2}{2}}e^{\alpha_{k,s} \hat{a}^\dagger_s}\lvert 0\rangle$, $s=l,r$, and $\lvert k;\alpha_{k,l},\alpha_{k,r}\rangle = \lvert k\rangle\otimes \lvert\alpha_{k,r}\rangle\otimes \lvert\alpha_{k,l}\rangle$. Substituting the ansatz, (\ref{eq:strn:me02}) into (\ref{eq:strn:me01}) produces two states of light $\lvert \alpha_{k,s}\rangle$ and $\hat{a}^\dagger \lvert\alpha_{k;l,s}\rangle$ that are not necessarily orthogonal. Decomposing  the state $\hat{a}^\dagger \lvert\alpha_{k,s}\rangle $ into its orthogonal form  $\hat{a}^\dagger \lvert\alpha_{k,s}\rangle = \lvert \alpha_{\perp,k,s}\rangle + \alpha^*_{k,s}\lvert\alpha_{k,s}\rangle$ where $\langle \alpha_{\perp,k,s} \lvert\alpha_{k,s}\rangle =0$~\citep{garcia-ripoll2005,hussain2014}  gives the equations for the evolution of  $\alpha_{k',s}$ and $\alpha_{k,s}$ and they read 
\begin{align}
\label{eq:strn:me03}
\frac{d\alpha_{k,l}}{dt }&= - \frac{ig(2k - N)}{2}\alpha_{k,l},\nonumber\\ 
%\frac{d\alpha^*_{k',l}}{dt}& =  \frac{ig(2k' - N)}{2}\alpha^*_{k',l},\\
\frac{d\alpha_{k,r}}{dt }&=  \frac{ig(2k - N)}{2}\alpha_{k,r}, 
%\frac{d\alpha^*_{k',r}}{dt}& = -\frac{ig(2k' - N)}{2}\alpha^*_{k',r}\nonumber,
\end{align}
whose solutions are 
\begin{align}
\label{eq:papc04}
\alpha_{k,l}(t) & = \alpha_{l} e^{- i(2k-N)\tfrac{gt}{2}},\nonumber\\
\alpha_{k,r}(t) & = \alpha_{r} e^{ i(2k-N)\tfrac{gt}{2}}.
%\alpha^*_{k',l}(t) & =  \alpha^*_{0,l} e^{ i(2k'-N)\tfrac{gt}{2}},\nonumber\\
%\alpha^*_{k',r}(t) & = \alpha^*_{0,r} e^{- i(2k'-N)\tfrac{gt}{2}}.
\end{align}
Clearly we see that the effect of the atom-light interaction is to impact a phase on the photons. This phase carrying information can be extracted in an interference measurement. 

Eliminating the $\alpha_{k,s}$ in the evolution of $\rho_{kk'}$ using~(\ref{eq:strn:me03}) and their solutions~(\ref{eq:papc04})~\cite{garcia-ripoll2005,hussain2014} give the following equation for the matrix elements $\rho_{kk'}$
\begin{align}
\label{eq:strn:me04}
\frac{d\rho_{mm'}}{dt} & =  i\Omega\frac{\sqrt{m(N-m+1)}}{2} \langle\alpha_m\lvert\alpha_{m-1}\rangle\rho_{m-1,m'}\nonumber\\
& + i\Omega\frac{\sqrt{(m+1)(N-m)}}{2} \langle\alpha_m\lvert\alpha_{m+1}\rangle\rho_{m+1,m'}\nonumber\\
& - i\Omega\frac{\sqrt{m'(N-m'+1)}}{2} \langle\alpha_{m'-1}\lvert\alpha_{m'}\rangle\rho_{m,m'-1}\nonumber\\
& - i\Omega\frac{\sqrt{(m'+1)(N-m')}}{2} \langle\alpha_{m'+1}\lvert\alpha_{m'}\rangle\rho_{m,m'+1}\nonumber\\
& - \gamma(m - m')\rho_{mm'},
\end{align}
where $\langle \alpha_m\lvert\alpha_{m\pm1}\rangle$ is the overlap of light coherent state that interacted with $m$ atoms and light coherent state that interacted with $m \pm 1$, respectively,
\begin{align}
\label{eq:strn:me05}
\langle\alpha_{m}\lvert\alpha_{m+1}\rangle &= e^{-(|\alpha_{l}|^2+|\alpha_{r}|^2)}e^{|\alpha_{l}|^2e^{-igt}}e^{|\alpha_{r}|^2e^{igt}},\nonumber \\
\langle\alpha_{m}\lvert\alpha_{m-1}\rangle &= e^{-(|\alpha_{l}|^2+|\alpha_{r}|^2)}e^{|\alpha_{l}|^2e^{igt}} e^{|\alpha_{r}|^2e^{-igt}}.
%\langle\alpha_{m'+1}\lvert \alpha_{m'}\rangle &=  e^{-(|\alpha_{0,l}|^2+|\alpha_{0,r}|^2)} e^{|\alpha_{0,l}|^2e^{igt}} e^{|\alpha_{0,r}|^2e^{-igt}}\nonumber,\\
%\langle\alpha_{m'-1}\lvert\alpha_{m'}\rangle &= e^{-(|\alpha_{0,l}|^2+|\alpha_{0,r}|^2)} e^{|\alpha_{0,l}|^2e^{-igt}} e^{|\alpha_{0,r}|^2e^{igt}}\nonumber.
\end{align}
Equation~(\ref{eq:strn:me04}) is a time dependent differential equation as determined by the phases (\ref{eq:strn:me05}). 

At the beamsplitter, photons that left each well are recombined as shown in Fig.~\ref{fig1} and sorted into bins $c$ and $d$ according to (\ref{eq:strn:me07}). The state after recombination at the beamsplitter becomes
\begin{align}
\label{eq:strn:me08}
\rho_\mathrm{BS}(t) & = \sum_{k,k'} \rho_{kk'}(t) e^{-\left(\frac{\lvert \alpha_{k,l}\rvert^2}{2}+\frac{\lvert \alpha_{k,r}\rvert^2}{2}\right)}e^{\frac{\alpha_{k,l} +i\alpha_{k,r}}{\sqrt{2}}\hat{a}^\dagger_c } \nonumber\\
&\times e^{\frac{i\alpha_{k,l} +\alpha_{k,r}}{\sqrt{2}}\hat{a}^\dagger_d } \lvert k,0,0\rangle \langle k', 0, 0\rvert e^{-\left(\frac{\lvert \alpha_{k',l}\rvert^2}{2}+\frac{\lvert \alpha_{k',r}\rvert^2}{2}\right)}\nonumber\\
& \times e^{\frac{\alpha^*_{k',l} -i\alpha^*_{k',r}}{\sqrt{2}}\hat{a}_c } e^{\frac{-i\alpha^*_{k',l} +\alpha^*_{k',r}}{\sqrt{2}}\hat{a}_d }. 
\end{align}
The probability of counting $n_c$ and $n_d$ photons in the bins $c$ and $d$, respectively, after recombination at the beamsplitter is obtained by the expectation of the projection operator $\mathcal{P}=\lvert n_c, n_d\rangle\langle n_c, n_d\rvert$ in the atoms subspace,
\begin{align}
\label{eq:strn:me09}
P(n_c, n_d) & = \mathrm{Tr}[\rho_\mathrm{BS}(t)\mathcal{P}],\nonumber\\
& = \frac{1}{n_c!}\frac{1}{n_d!}\sum_{k}\rho_{kk}(t)e^{-\left(\lvert\alpha_{k,l}\rvert^2 + \lvert\alpha_{k,r}\rvert^2\right)}\nonumber\\ &\times\left(\frac{\lvert\alpha_{k,l} \rvert^2 +i\alpha_{k,r}\alpha^*_{k,l} - i\alpha_{k,l}\alpha^*_{k,r} + \lvert\alpha_{k,r}\rvert^2 }{2}\right)^{n_c}\nonumber\\
&\times\left(\frac{\lvert\alpha_{k,l} \rvert^2 -i\alpha_{k,r}\alpha^*_{k,l} +i\alpha_{k,l}\alpha^*_{k,r} + \lvert\alpha_{k,r}\rvert^2 }{2}\right)^{n_d}.
\end{align}
Thus the state of the atoms given that $n_c$ and $n_d$ photons have been detected becomes
\begin{align}
\label{eq:strn:me10}
\rho_{n_c,n_d}(t) &= \frac{1}{n_c!}\frac{1}{n_d!}\frac{1}{P(n_c,n_d)}\sum_{k,k'}\rho_{kk'}(t)\times\nonumber\\ &e^{-\left(\frac{\lvert\alpha_{k,l}\rvert^2}{2} + \frac{\lvert\alpha_{k',l}\rvert^2}{2} + \frac{\lvert\alpha_{k,r}\rvert^2}{2} \frac{\lvert\alpha_{k',r}\rvert^2}{2} \right)}\nonumber\\
&\times\left(\frac{\alpha_{k,l}\alpha^*_{k',l} +i\alpha_{k,r}\alpha^*_{k',l} - i\alpha_{k,l}\alpha^*_{k',r} + \alpha_{k,r}\alpha^*_{k',r} }{2}\right)^{n_c}\nonumber\\
&\times\left(\frac{\alpha_{k,l}\alpha^*_{k',l} -i\alpha_{k,r}\alpha^*_{k',l} + i\alpha_{k,l}\alpha^*_{k',r} + \alpha_{k,r}\alpha^*_{k',r} }{2}\right)^{n_d}\nonumber\\
&\times\lvert k,N-k\rangle\langle k',N-k'\rvert.
\end{align}
%

%==========================================
\section{Most Probable Outcome\label{sec:photonprobability} }
%==========================================
Let us first consider that $gt=0$, which occurs when there are no atom-light interactions. In this limit, the amplitudes $A_{n_c.n_d}(k)$ can be taken outside the sum over $k$ and the contribution of the atoms to the probability (\ref{eq:sm03}) sums up to unity, $\sum_k \lvert C_k \rvert^2 = 1$. Then, the probability (\ref{eq:sm03}) of obtaining $n_c$ and $n_d$ photons after measurement simplifies $P(n_c,n_d) = \sum_k \lvert C_k\rvert^2\lvert A_{n_c,n_d}(k)\rvert^2 = \lvert A_{n_c,n_d}(k)\rvert^2$. Using  (\ref{eq:sm04}) and (\ref{eq:sm05}), 
\begin{align}
\label{eq:sma06a}
\lvert A_{n_c,n_d}(k)\rvert^2 =& e^{-\lvert \alpha_l\rvert^2 -\lvert\alpha_r\rvert^2} \frac{\left(\dfrac{\lvert \alpha_l\rvert^2 +\lvert\alpha_r\rvert^2}{2}\right)^{n_c}}{n_c!}\nonumber\\
\times&  \frac{\left(\dfrac{\lvert\alpha_l\rvert^2 + \lvert\alpha_r\rvert^2}{2}\right)^{n_d}}{n_d!}.
\end{align}
Observe that $\lvert A_{n_c,n_d}(k)\rvert^2$ can be written as a product of two probabilities $P(n_c)$ and $P(n_d)$ as $\lvert A_{n_c,n_d}(k)\rvert^2 = P(n_c)P(n_d)$ where 
\begin{equation}
\label{eq:sma07}
P(n_j) = e^{-\frac{|\alpha_l|^2+ |\alpha_r|^2}{2}}\left(\frac{|\alpha_l|^2+ |\alpha_r|^2}{2}\right)^{n_j}\frac{1}{n_j!},
\end{equation}
$j =c,d$. Suppose that $n_c$, $n_d \gg 1$ and using Stirling's approximation, one can write $P(n_j)$ as 
\begin{equation}
\label{eq:sma10}
P(n_j) = \frac{1}{\sqrt{\pi \left(\lvert \alpha_r\rvert^2 + \lvert\alpha_l\rvert^2 \right)}}\exp\left[-\frac{\left(n_j - \frac{\alpha_r\rvert^2 - \lvert\alpha_l\lvert^2}{2} \right)^2}{\lvert \alpha_r\rvert^2 + \lvert\alpha_l\rvert^2}\right].
\end{equation} 
It is then clear that the most probable outcome occurs for $n_d = n_c = (\lvert \alpha_r\rvert^2 + \lvert\alpha_l\rvert^2)/2$, see also Refs.~\cite{ilo-okeke2014,ilo-okeke2016} that proved this using a different method. 

For the number of photons $n_c,n_d>1$ and using Stirling's approximation, the magnitude of the function $A_{n_c,n_d}(k)$ becomes
\begin{align}
\label{eq:sma02}
|A_{n_c,n_d}(k)| =& \left(\frac{\lvert\alpha_l\rvert^2 + \lvert\alpha_r\rvert^2}{n_c + n_d}\right)^{\frac{n_c + n_d}{2}} e^{\frac{n_c + n_d -\lvert\alpha_l\rvert^2 - \lvert\alpha_r\rvert^2}{2}}\nonumber \\
\times &\left(\frac{1}{4\pi^2 n_c n_d}\right)^{1/4} e^{-\frac{X_0}{4}(k- N/2 - x_0)^2},
\end{align}
where
\begin{align}
\label{eq:sm09a}
x_0 &= \frac{1}{2gt}\left(\phi - \arcsin\left[ \frac{|\alpha_l|^2 + |\alpha_r|^2}{2|\alpha_l\alpha_r|}\frac{n_c - n_d}{n_d + n_c}\right]\right),\\
\label{eq:sm10a}
X_0& = (gt)^2\left(\frac{n_c + n_d}{n_cn_d}\right)\Bigg[(n_c + n_d)^2\left(\frac{2|\alpha_l\alpha_r|}{|\alpha_l|^2 + |\alpha_r|^2}\right)^2\nonumber\\ 
&- (n_d - n_c)^2 \Bigg],
\end{align}
and $\phi =  \phi_l - \phi_r$ is the relative phase between the coherent light beams on the left and right as shown in Fig.~\ref{fig1}.  

%==========================================
\section{The Approximate State of BEC After Photon Detection\label{sec:approximatestate}}
%==========================================
Working in the \emph{x}-basis, the probability amplitude of the initial state $\lvert \alpha,\beta\rangle$ (\ref{eq:strn07})
\begin{equation}
\label{eq:sm07a}
C_k = \sqrt{\frac{N!}{k!(N-k)!}}\left(\eta_l\right)^{k}\left(\eta_r\right)^{N-k},
\end{equation}
where $\eta_l = (\alpha + \beta)/\sqrt{2}$ and $\eta_r = (\beta - \alpha)/\sqrt{2}$. For $N\gg1$, the dominant contribution to $C_k$ come from terms around $k =N\lvert\eta_l\rvert^2$ that has a width $\Delta k \sim \sqrt{N}$. Using Stirling's approximation, the magnitude of $C_k$  becomes
\begin{equation}
\label{eq:sm07b}
\lvert C_k\rvert \approx \left(\frac{1}{2\pi N\lvert\eta_l\eta_r\rvert^2}\right)^{1/4} e^{-\frac{1}{4N\lvert\eta_l\rvert^2\lvert\eta_r\rvert^2}\left(k - N\lvert\eta_l\rvert^2\right)^2}. 
\end{equation}
Hence the initial probability distribution $\lvert C_k \rvert^2$ of the atoms in the state $\lvert \alpha,\beta\rangle$ can be approximated to  Gaussian.

The probability density $P(n_c,n_d)$ of detecting $n_c, n_d$ photons in a measurement~(\ref{eq:sm03}) is easily calculated using $\lvert C_k\rvert$ (\ref{eq:sm07b}), and $\lvert A_{n_c,n_d}(k)\rvert$ (\ref{eq:sma02}), and it evaluates to 
\begin{align}
\label{eq:sma03}
P(n_c,n_d) =& \sqrt{\frac{1}{4\pi^2 n_cn_d(1 + N\lvert\eta_l\eta_r\rvert^2X_0)}}\nonumber\\
\times&\left(\frac{\lvert\alpha_l\rvert^2 + \lvert\alpha_r\rvert^2}{n_c + n_d}\right)^{n_c + n_d} e^{n_c + n_d -\lvert\alpha_l\rvert^2 - \lvert\alpha_r\rvert^2}\nonumber\\
\times& \exp\left[-\frac{X_0\left(x_0 - \frac{N(\lvert\eta_l\rvert^2 - \lvert\eta_r\rvert^2)}{2}\right)^2}{2(1 + N\lvert\eta_l\eta_r\rvert^2X_0)}\right].
\end{align}

The approximate state of BEC after detecting $n_c$ and $n_d$ photons using (\ref{eq:sm07b}), (\ref{eq:sma02}), and (\ref{eq:sma03})  in  (\ref{eq:sm06}) becomes
\begin{align}
\label{eq:sma04}
\lvert\psi^\mathrm{approx}_{n_c,n_d}\rangle \approx& \left(\frac{1 + N\lvert\eta_l\eta_r\rvert^2 X_0 }{2\pi N\lvert\eta_l\eta_r\rvert^2 }\right)^{1/4}\nonumber\\
\times& \sum_k \exp\Bigg[-\frac{N|\eta_l\eta_r|^2 X_0 + 1}{4N|\eta_l\eta_r|^2}\nonumber\\
\times& \Bigg(k -   \frac{N|\eta_l|^2 + N|\eta_l\eta_r|^2 X_0 (N/2 + x_0)}{N|\eta_l\eta_r|^2 X_0 + 1}\Bigg)^2\Bigg]\nonumber\\
\times& e^{ik\varphi}e^{i(n_c\phi_c(k) + n_d\phi_d(k))}\lvert k, N-k\rangle,
\end{align}
where $\varphi$ is the initial relative phase of the BEC in left and right well, $\phi_c(k)$ is the argument of $\alpha_c(k)$, and $\phi_d(k)$ is the argument of $\alpha_d(k)$. It immediately follows that
\begin{align}
\label{eq:sm08}
P(k\lvert n_c,n_d) & = \sqrt{\frac{N|\eta_l\eta_r|^2X_0 + 1}{2\pi N|\eta_l\eta_r|^2}}\exp\Bigg[-\frac{N|\eta_l\eta_r|^2 X_0 + 1}{2N|\eta_l\eta_r|^2}\nonumber\\
& \times \Bigg(k -  \frac{N|\eta_l|^2 + N|\eta_l\eta_r|^2 X_0 (N/2 + x_0)}{N|\eta_l\eta_r|^2 X_0 + 1}\Bigg)^2\Bigg]. 
\end{align} 
In order to simplify the expressions, we will work about the most probable outcome, with $n_d = n_c$. We will use balanced detection scheme $\alpha_l = \alpha_r$ (i.e. there is no relative phase between the light beams). At these values, $x_0=0$ and $X_0 = 8g^2t^2n_c$. Substituting these  values in (\ref{eq:sm07}), (\ref{eq:sma04}) and (\ref{eq:sm08}) give the expression (\ref{eq:sma08})  in the main text.

%\bibliographystyle{apalike}% style could be plain i.e.\bibliographystyle{plain,unsrt,amsplain,apsrev,agsm,apalike}. However this style sorts it in the form as they are cited. apalike is APA, and agsm is Harvard
%\bibliographystyle{apsrev}
%\bibliography{ReferenceFile}%The following command as it is allows me to have a central .bib file which I can always update

%
%
\end{document}